\documentclass[fleqn,usenatbib]{mnras}

\usepackage{newtxtext,newtxmath}

\usepackage[T1]{fontenc}

\DeclareRobustCommand{\VAN}[3]{#2}
\let\VANthebibliography\thebibliography
\def\thebibliography{\DeclareRobustCommand{\VAN}[3]{##3}\VANthebibliography}

\usepackage{graphicx}	
\usepackage{amsmath}	

\makeatletter
\newcommand\sendemail[3]{
\edef\@tempa{mailto:#1?subject=#2 }%
\edef\@tempb{\expandafter\html@spaces\@tempa\@empty}%
\href{\@tempb}{#3}}

\catcode\%=11
\def\html@spaces#1 #2{#1
\catcode\%=14
\makeatother

\title{The distribution of stellar orbits in {\sc Eagle} galaxies - the effect of mergers, gas accretion, and secular evolution}

\author[G. Santucci et al.]{\parbox{\textwidth}{
Giulia Santucci,$^{1,2}$\thanks{E-mail: giulia.santucci@uwa.edu.au}
Claudia Del P. Lagos,$^{1,2}$
Katherine E. Harborne,$^{1,2}$
Aaron Ludlow,$^{1,2}$
Caro Foster,$^{3,2}$
Richard McDermid, $^{4,2}$
Adriano Poci, $^{5}$
Katy L. Proctor$^{1,2}$
Sabine Thater,$^{6}$
Glenn van de Ven,$^{6}$
Ling Zhu, $^{7}$
Daniel Walo Mart\'{i}n
\\}
\vspace{0.4cm}
\\
\parbox{\textwidth}{
$^{1}$ International Centre for Radio Astronomy Research (ICRAR), M468, University of Western Australia, 35 Stirling Hwy, Crawley, WA 6009, Australia\\
$^{2}$ARC Centre of Excellence for All Sky Astrophysics in 3 Dimensions (ASTRO 3D), Australia\\
$^{3}$School of Physics, University of New South Wales, Sydney, NSW 2052, Australia\\
$^{4}$Department of Physics and Astronomy, Macquarie University, Sydney, NSW 2109, Australia\\
$^{5}$ Centre for Extragalactic Astronomy, University of Durham, Stockton Road, Durham DH1 3LE, United Kingdom\\
$^{6}$ Department of Astrophysics, University of Vienna, Türkenschanzstrasse 17, 1180 Vienna, Austria\\
$^{7}$Shanghai Astronomical Observatory, Chinese Academy of Sciences, 80 Nandan Road, Shanghai 200030, China\\
}
}

\date{Accepted XXX. Received YYY; in original form ZZZ}

\pubyear{2023}

\begin{document}
\label{firstpage}
\pagerange{\pageref{firstpage}--\pageref{lastpage}}
\maketitle

\begin{abstract}
The merger history of a galaxy is thought to be one of the major factors determining its internal dynamics, with galaxies having undergone different types or mergers (e.g. dry, minor or major mergers) predicted to show different dynamical properties. We study the instantaneous orbital distribution of galaxies in the {\sc Eagle} simulation, colouring the orbits of the stellar particles by their stellar age, in order to understand whether stars form in particular orbits (e.g. in a thin or thick disc). We first show that {\sc Eagle} reproduces well the observed stellar mass fractions in different stellar orbital families as a function of stellar mass and spin parameter at \(z=0\). We find that the youngest stars reside in a thin disc component that can extend to the very inner regions of galaxies, and that older stars have warmer orbits, with the oldest ones showing orbits consistent with both hot and counter-rotating classifications, which is consistent with the trend found in the Milky-Way and other disc galaxies. We also show that counter-rotating orbits trace galaxy mergers - in particular dry mergers, and that in the absence of mergers, counter-rotating orbits can also be born from highly misaligned gas accretion that leads to star formation.

\end{abstract}




\section{Introduction}

A galaxy's assembly history is thought to be one of the major factors determining its internal kinematic structure \citep[e.g.,][]{White1979, Fall1980, Park2019}. It is thus expected that the internal orbital structure of a galaxy should contain important clues about a galaxy's past. In particular, stellar dynamics provide a fossil record of the formation history of galaxies. Stars that were born and remain in quiescent environments tend to be on regular rotation-dominated orbits, so their galaxies are likely dominated by fast-rotating, flat stellar discs with recent star formation (e.g., \citealt{Fall1980,Lagos2017}). On the other hand, stars born from turbulent gas or that have been dynamically heated after birth will be on warmer orbits with more random motions (e.g. \citealt{Leaman2017}). Dynamical heating mechanisms that can affect stars include major mergers (e.g. \citealt{Benson2004, House2011, Helmi2012, Few2012, Ruizlara2016,Lagos2018a}) and long-term secular heating of the disc via internal instabilities (e.g. \citealt{Jenkins1990,Aumer2016,Grand2016}), for example. Major gas-poor mergers, in particular, can play a key role in destroying stellar discs and in creating dynamically hot spheroidal components in the remnant galaxies \citep{Barnes1996,Cox2006, Hoffman2010, Bois2010, Bois2011, Naab2014, Pillepich2015,Lagos2018a}.

Stellar kinematics are also expected to be systematically correlated with stellar ages (e.g. \citealt{vandeSande2018,Trayford2019}). Observations have found that old stars dominate the light of pressure-supported bulges, while younger stars live on thinner discs. In general, stars born together tend to live on similar orbits (e.g., \citealt{Bird2013, Stinson2013}). If we study the phase-space distribution of stars or stellar components in a galaxy, we can potentially infer information on whether these components have formed in the galaxy itself or are the results of merging events. It is expected that both merging events and star formation episodes play an important role in setting a galaxy's structure and its stellar populations. Thus, the more information we can gather about the different orbital structures in a galaxy, the closer we can get to unveiling its evolutionary history.

Dynamical models can be used to infer the orbital distributions within galaxies. One way to parametrise this distribution is through the `circularity plane'; the distribution of orbits in the \(\lambda_z-R\) plane, where \(\lambda_z\) represents the $z$-axis component of the angular momentum of the stars normalised by the maximum angular momentum a star with the same binding energy could have (i.e. that of a circular orbit), and the time-averaged radius, $r$, as a proxy for the binding energy. The distribution of stellar orbits in a galaxy can thus be represented by the probability density of orbits in the phase-space of  $\lambda_{\rm z}$ versus $r$. The different dynamical structures of galaxies can also be identified within this phase-space plane with, for example, the subdivision of the stellar component in a galaxy into cold ($\lambda_{\rm z} > 0.8$), warm ($0.25 < \lambda_{\rm z} < 0.8$), hot ($| \lambda_{\rm z}| < 0.25$), and counter-rotating
($\lambda_{\rm z} < - 0.25$) orbits (e.g., \citealt{Zhu2018nature, Zhu2018hubble, Jin2019, Jin2020, Poci2021, Poci2022, Santucci2022}).

In recent years, Integral Field Spectroscopy (IFS) surveys such as SAURON (Spectroscopic Areal Unit for Research on Optical Nebulae; \citealt{deZeeuw2002}), ATLAS\(^{\rm 3D}\) \citep{Cappellari2011}, CALIFA (Calar Alto Legacy Integral Field Array survey; \citealt{Sanchez2012}), SAMI (Sydney-Australian-Astronomical-Observatory Multi-object Integral-Field Spectrograph) Galaxy Survey \citep{Croom2012, Bryant2015,Croom2021}, MASSIVE \citep{Ma2014}, MaNGA (Mapping Nearby Galaxies at Apache Point Observatory; \citealt{Bundy2015}) and the Fornax 3D survey \citep{Sarzi2018} have provided us with rich resolved views of galaxies, allowing their structure and evolution to be investigated in detail through the mapping of stellar kinematics across individual galaxies. These IFS surveys have made possible the use of techniques such as the Schwarzschild orbit-superposition method \citep{Schwarzschild1979} to estimate the internal mass distribution, intrinsic stellar shapes, and orbit distributions of galaxies across the Hubble sequence. The stellar orbit circularity distribution within 1 effective radius ($R_{\rm e}$, radius of the isophote containing half of the total luminosity;  \citealt{deVaucouleurs1948}) has been obtained for 300 CALIFA galaxies across the Hubble sequence \citep{Zhu2018nature, Zhu2018hubble} and for a few hundred early-type galaxies (ETGs) in MaNGA \citep{Jin2020}, SAMI \citep{Santucci2022}, and ATLAS$^{\rm 3D}$ \citep{Thater2023}. These works show that the fraction of dynamically hot orbits increases from low-mass to high-mass galaxies in the local Universe, consistently with an increase in the mass fraction of a classical bulge (i.e. those with triaxial shape, high S\'ersic index, and dominated by velocity dispersion; \citealt{Weinzirl2009}). However, even though the studies of galaxies' orbital components in large IFS surveys have already started, the formation of these orbital components has not yet been quantitatively connected to the merger history of galaxies.

The details of merger histories have been (partially) uncovered in the nearest galaxies, where either single stars in the stellar halos can be resolved, for example in the case of the Milky Way \citep{Helmi2018,Belokurov2018, Helmi2020}, the Andromeda galaxy \citep{Dsouza2018,Dsouza2018Nat}, and NGC 5128 \citep{Rejkuba2011}, or where individual globular clusters in the halos are detected \citep{Forbes2016, Beasley2018, Kruijssen2019}, or where streams have been detected \citep{Merritt2016,Harmsen2017}. Recently, studies have attempted to constrain the global ex-situ fractions - the amount of stellar mass that formed in galaxies other than the main galaxy studied \citep{Boecker2020,Spavone2020, Davison2021, Angeloudi2023}, or the typical mass of the accreted satellites \citep{Pinna2019, Martig2021} combining stellar populations and orbital distributions in edge-on lenticular galaxies obtained from IFS data \citep{Poci2019,Poci2021}. However, so far the assembly and merger histories of most galaxies remain hidden, including in the nearby Universe.

Cosmological hydrodynamical simulations make it possible to trace back the formation of galactic structures in great detail \citep{Lagos2018,Obreja2019, Trayford2019,Du2021,Pulsoni2021, Lagos2022, Proctor2023}, including adopting dynamical decomposition methods similar to those applied to observational data \citep{Xu2019}, finding quantitatively consistent results between simulations and observations. The number of studies connecting the dynamical structures of galaxies with their accretion histories is increasing. It is then essential to exploit the additional information we can derive from simulated galaxies in order to connect galaxy structures with their formation and evolution. Having a consistent method to measure dynamical structures will allow us not only to directly compare across simulation outputs and observations, but to also shed light on the connection between the orbital properties and the assembly histories of galaxies. For example, \cite{Zhu2022halo} focused on the orbital distributions of galaxies in the IllustrisTNG \citep{Pillepich2018, Springel2018, Nelson2019} and {\sc Eagle} \citep{Schaye2015, Crain2015} simulations, identifying hot inner stellar halo components in a way that is also consistent and achievable with deep IFS observations. They show that IFS observations and dynamical models of the inner regions of galaxies can provide a way to quantitatively determine the mass and time of ancient massive mergers \citep{Zhu2022fornax}.

In this paper we build on these studies and explore the effect of galaxy mergers on the orbital structure of galaxies that can be obtained for many galaxies with large IFS surveys, in order to understand whether stars form on particular orbits (e.g. in a thin or thick disc) or whether they migrate to their $z\sim 0$ configuration. We employ the {\sc Eagle} simulations, as these have been shown to successfully produce galactic structures, such as bulge sizes, that are similar to observations in galaxies with $M_{\star} \ge 10^{10} M_{\odot}$ \citep{Lange2016}, as well as to reproduce the fundamental plane of elliptical galaxies \citep{deGraaff2023} and the diversity of stellar rotation-to-pressure support observed in local Universe galaxies \citep{Lagos2018,vandeSande2019}. We also explore how well {\sc Eagle} reproduces current observational constraints on the orbital structures of galaxies from the local Universe. 

This paper is organised as follows. In Section 2 we describe the sample of galaxies and the galaxy properties we use in this analysis. Section 3 presents the orbital fractions we derive for the {\sc Eagle} sample, comparing them with results from observations. The connection between the orbital fractions and a galaxy's evolution history is discussed in Section 4. Our conclusions are given in Section 5.

\section{Data}
\subsection{The {\sc Eagle} simulation}

The {\sc Eagle} simulation suite (\citealt{Schaye2015}, hereafter S15;  \citealt{Crain2015}) consists of a large number of cosmological hydrodynamic simulations with different resolutions, volumes, and subgrid physics models, adopting the \cite{Plank2014} cosmology. S15 introduced a reference model, within which the parameters of the sub-grid models governing energy feedback from stars and accreting black holes (BHs) were calibrated to ensure a good match to the $z = 0.1$ galaxy stellar mass function, the size-mass relation of present-day star-forming galaxies and the black hole-stellar mass relation (see \citealt{Crain2015} for details). In this work, we take the largest volume with the highest resolution presented in S15. Table~\ref{TableSimus} summarises the parameters of this simulation.

\begin{table}
\begin{center}
  \caption{Specifications of the {\sc Eagle} Ref-L100N1504 simulation used in this paper. The rows list:
    (1) initial particle masses of gas and (2) dark
    matter, (3) comoving Plummer-equivalent gravitational
    softening length, and (4) maximum physical
    gravitational softening length. Units are indicated in each row. {\sc Eagle}
    adopts (3) as the softening length at $z\ge 2.8$, and (4) at $z<2.8$. This simulation
    has a side length of $L=100$~$\rm cMpc^3$. Here, pkpc and ckpc refer to proper and comoving kpc, respectively. }\label{TableSimus}
\begin{tabular}{l l l l}
\\[3pt]
\hline
& Property & Units & Value \\
\hline
(1)& gas particle mass & $[\rm M_{\odot}]$ & $1.81\times 10^6$\\
(2)& dark matter (DM) particle mass & $[\rm M_{\odot}]$ & $9.7\times 10^6$\\
(3)& Softening length & $[\rm ckpc]$ & $2.66$\\
(4)& max. gravitational softening & $[\rm pkpc]$& $0.7$ \\
\end{tabular}
\end{center}
\end{table}

A major aspect of the {\sc Eagle} project is the use of state-of-the-art sub-grid models that capture unresolved physics (i.e. processes happening below the resolution limit). The sub-grid physics modules adopted by {\sc Eagle} are: (i) radiative cooling and photoheating \citep{Wiersma2009}, (ii) star formation \citep{Schaye2008}, (iii) stellar evolution and chemical enrichment \citep{Wiersma2009b}, (iv) stellar feedback \citep{DallaVecchia2012}, and (v) BH growth and active galactic nucleus (AGN) feedback 
\citep{Rosas-Guevara2015}. 

The {\sc Eagle} simulations were performed using an extensively modified version of the parallel N-body smoothed particle hydrodynamics (SPH) code GADGET-3 \citep{Springel2005Gadget}. Among those modifications are updates to the SPH technique, which are collectively referred to as ``Anarchy'' (see \citealt{Schaller2015} for an analysis of the impact of these changes on the properties of simulated galaxies compared to standard SPH). We use SUBFIND \citep{Springel2001,Dolag2009} to identify self-bound overdensities of particles within halos (i.e. substructures). These substructures are the galaxies in {\sc Eagle}.

\subsection{Merger Trees}

We identify mergers using the merger trees available in the {\sc Eagle} database \citep{McAlpine2016}. These merger trees were created using the {\sc D - Trees} algorithm of \cite{Jiang2014}. \cite{Qu2017} described how this algorithm was adapted to work with {\sc Eagle} outputs. Galaxies that went through mergers have more than one progenitor, and for our purpose, we track the most massive progenitors of merged galaxies and define that as the main branch. The trees stored in the public database of {\sc Eagle} connect 29 epochs. The time span between snapshots can range from $\approx$ 0.3 Gyr to $\approx$ 1 Gyr. We use these snapshots to analyse the evolution of the orbital components in galaxies and the effect of galaxy mergers.

\subsection{Stellar mass, luminosities, and the half-mass radius}

The stellar mass of {\sc Eagle} galaxies is measured using a spherical aperture. This gives similar results to the 2D Petrosian aperture used in observational works, and provides an orientation-independent mass measurement for each galaxy. Following previous studies based on the {\sc Eagle} simulations, we take as stellar mass the mass of the stellar particles within a 30~pkpc (pkpc refers to proper kpc) aperture centred at the centre of potential (e.g. \citealt{Schaye2015}). 

We compute the projected stellar half-mass radius for each galaxy as the projected physical radius enclosing half of the stellar mass in the subhalo, averaged over three orthogonal projections (referred to as $r_{\rm 50}$ hereafter).

To compute the \(r\)-band light, we model each stellar particle as a single stellar population (SSP) with the age and metallicity of the particle, using the EMILES population synthesis models \citep{Vazdekis2016} and adopting a \cite{Chabrier2003} initial mass function. This provides a spectrum for each stellar particle. The r-band light is then the result of the convolution of the spectra with the Sloan Digital Sky Survey r-band filter \citep{Fukugita1996}. We use these r-band luminosities to calculate r-band luminosity weighted properties of galaxies, which are more directly comparable to observational results from IFS surveys (e.g. \citealt{Santucci2022,Santucci2023}).

\subsection{{\sc Eagle} Sample Selection}

We select galaxies from the {\sc Eagle} Ref-L100N1504 simulation, spanning a stellar mass range of $10^{9} - 10^{12}\,\rm  M_{\odot}$ at all snapshots between $z=0$ and $z=2$. We note that due to the resolution of the simulation, dynamical properties of galaxies are not reliable for stellar masses $\lesssim 10^{9.5}\,\rm M_{\odot}$ (see Appendix~A in \citealt{Lagos2017}, and \citealt{Ludlow2023}). Moreover, \cite{Ludlow2023} showed that a stellar mass $\gtrsim 10^{10}\,\rm M_{\odot}$ is needed for stellar kinematics to be resolved at the 3D half-mass radii of {\sc Eagle} galaxies; at lower masses, gravitational scattering between stellar and DM particles affects their kinematics and spatial distribution. It is important to note that DM mass resolution and gravitational scattering can also affect the circularity distributions of stellar disc particles by transforming galactic discs from flattened structures into rounder spheroidal systems, causing them to lose rotational support in the process \citep{Wilkinson2023}. We present a comparison between the {\sc Eagle} "Reference" simulation used here and the High-Resolution DM simulation from \cite{Ludlow2023} in Appendix \ref{app:hrdm}.
 
For completeness, we present our results down to stellar masses of $10^9\,\rm M_{\odot}$ (13199 galaxies), but we highlight the regime in which our results are likely affected by resolution. When looking at the connection between orbital fractions and mergers, we focus on galaxies with stellar masses $\gtrsim 10^{10}\,\rm M_{\odot}$ (3561 galaxies), since this is the regime where kinematic quantities can robustly be measured in {\sc Eagle}. 

During our analysis we often distinguish between galaxies that have or have not had mergers. To reliably trace minor and major mergers for all galaxies with stellar masses $>10^9\,\rm M_{\odot}$, we study all potential progenitors with stellar masses $\ge 10^8\,\rm M_{\odot}$. We make this distinction based on the merger history of galaxies from $z=2$ to $z=0$: galaxies with ``no mergers'' will be those that did not have a minor or major merger during that period, with minor and major mergers being those with a stellar mass ratio between the second most massive and most massive progenitors between $[0.1-0.3[$ and $[0.3,1]$, respectively. Note that we can trace mergers with mass ratios $<0.1$ but the sample would not be complete for all galaxies in our sample given the limitation of the progenitors needing to have a stellar mass $\ge 10^8\,\rm M_{\odot}$. Hence, we opt to only track minor and major mergers.
\begin{figure*}
\centering
\includegraphics[scale=0.29 ,trim=0cm 0cm 0cm 0cm, clip=True]{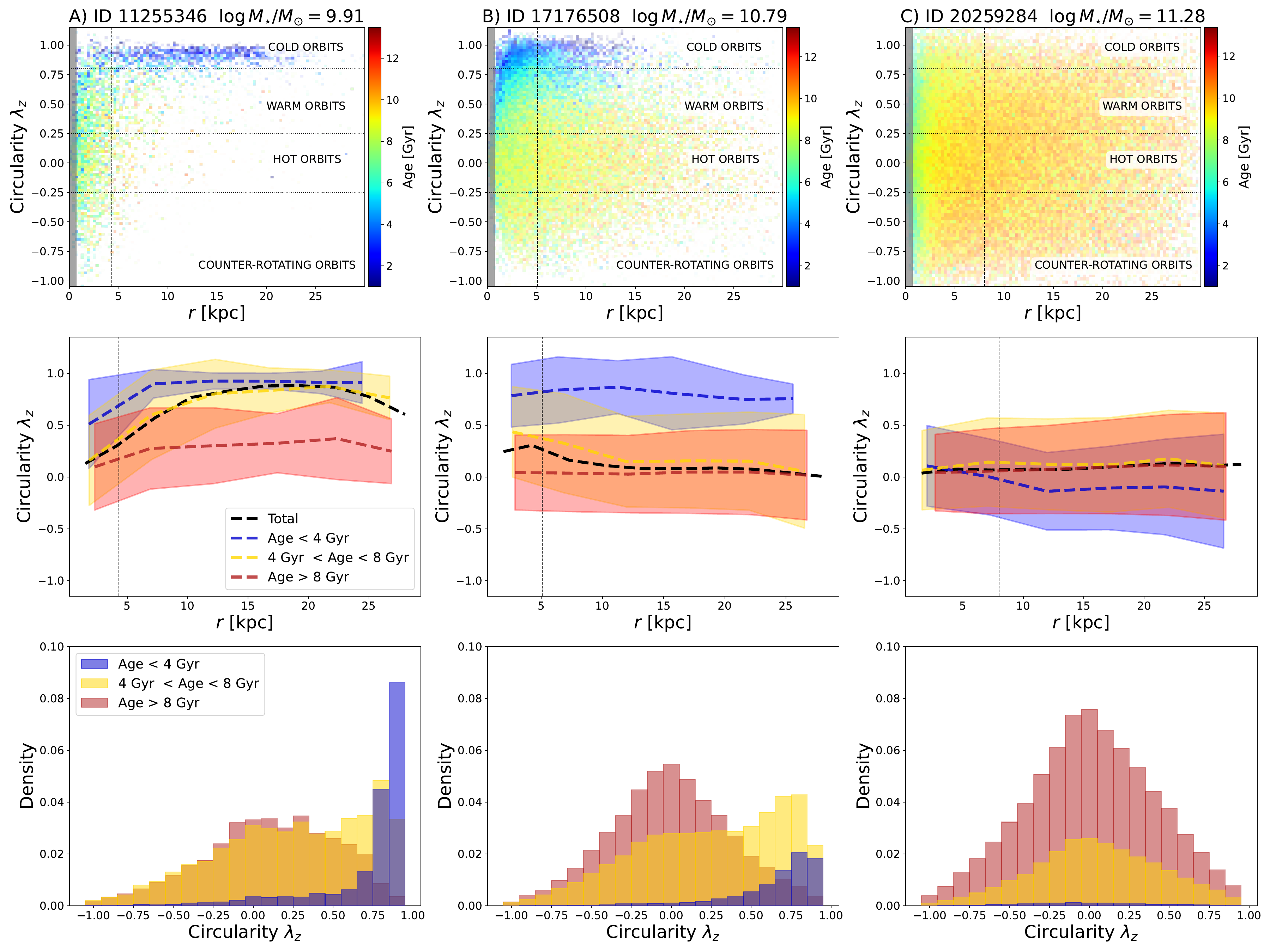}
\caption{Left column: Example Galaxy A, ID 11255346, with stellar mass $\log M_{\star}/\rm M_{\odot} \sim 9.91$. The top panel shows stellar particles in the $\lambda_{\rm z}$-radius plane, binned into cells. The colour represents the mass-weighted age, and the transparency reflects the total mass of each cell. The shaded grey area represents the region affected by the gravitational softening. The middle panel shows the average radial distribution of $\lambda_{\rm z}$ for all of the stellar particles (black line) and for stellar particles in three different age ranges (Age $< 4$ Gyr in blue, $4$ Gyr $<$ Age $< 8$ Gyr in yellow and Age $> 8$ Gyr in red) as solid lines. Shaded regions delimit the 1~$\sigma$ scatter around the median. The bottom panel shows the 1D distribution of $\lambda_{\rm z}$ for stellar particles in the same three different age ranges. The majority of the stellar particles in this galaxy are in the cold and warm components, with young stars residing primarily in the cold component. Older stars have a wider spread in their orbital distributions. Vertical dotted lines indicate 1~$r_{\rm 50}$. Middle column: Example Galaxy B, ID 17176508, with stellar mass $\log M_{\star}/\rm \rm M_{\odot} \sim 10.79$. The majority of the stellar particles in this galaxy are older than 4 Gyr, with a large fraction of the older particles ($> 8$ Gyr) in the hot component. Young stars have mostly cold orbits, and middle-aged particles have a wider spread in orbital distribution, with a peak contribution in the cold component. Right column: Example Galaxy C, ID 20259284, with stellar mass $\log M_{\star}/\rm \rm M_{\odot} \sim 11.28$. The majority of the stellar particles in this galaxy are old and located in the hot component. Middle-age particles show a similar distribution to the older stellar particles, with a peak in the hot component. }
\label{fig:example_1}
\end{figure*}

\subsection{Orbital decomposition of {\sc Eagle} galaxies}\label{sec:orbital_decomp}

To get to the orbital distribution and decomposition of stellar particles in {\sc Eagle} galaxies, we first calculate the ratio between the specific angular momentum of a stellar particle in the $z$ direction relative to the specific angular momentum of a circular orbit of the same energy, $\lambda_{\rm z}$ (a.k.a. circularity parameter). For this we follow the method introduced by \cite{Abadi2003} described below:

\begin{itemize}
\item We first define the $z$-axis of a galaxy as that parallel to the stellar angular momentum vector of the galaxy computed with all stellar particles of the subhalo with the spatial reference frame being centred on the position of the most bound particle, and the velocity reference being the mean mass-weighted velocity of all stellar particles in the subhalo.
\item Compute the specific angular momentum of each stellar particle in the $z$ direction, $j_{\rm z}$.
\item We then compute the potential and kinetic energy of all stellar particles in the subhalo and bin particles by their total energy, with bins containing 100 particles.
\item The specific angular momentum of a circular orbit with the total energy of a given stellar particle, $j_{\rm c}$, is then taken as the maximum $j_{\rm z}$ among the 100 particles in the bin of total energy of that stellar particle. 
\item We calculate a $\lambda_{\rm z}=j_{\rm z}/j_{\rm c}$ for each stellar particle.
\end{itemize}

With the calculation above, we construct a 2-dimensional distribution of $\lambda_{\rm z}$ vs projected distance (in the $\rm xy$ plane of the galaxy) for each galaxy. The top-left panel in Fig. \ref{fig:example_1} shows examples of these distributions for $3$ {\sc Eagle} galaxies at $z=0$, revealing the diversity of orbital structures that galaxies can have. We provide a detailed analysis of these example galaxies later in this section.

We separate orbits into four different components as follows: 
\begin{itemize}
    \item {\it cold orbits} selected as $\lambda_{\rm z} > 0.8$, are those with near-circular orbits;
    \item {\it warm orbits} selected as $0.25 < \lambda_{\rm z} < 0.8$, are somewhat disc-like, but with a considerable amount of angular momentum out of the disc ($z$) plane;
    \item {\it hot orbits} selected as $-0.25 < \lambda_{\rm z} < 0.25$, are those with near radial orbits;
    \item {\it counter-rotating orbits} are those with $\lambda_{\rm z} < -0.25$.
\end{itemize}
These thresholds were adopted to follow the analysis of \cite{Zhu2018nature, Jin2020, Santucci2022} and allow for easier comparison with observations.

\begin{figure*}
\centering
\includegraphics[scale=0.45 ,trim=0cm 0.5cm 0cm 0.4cm, clip=True]{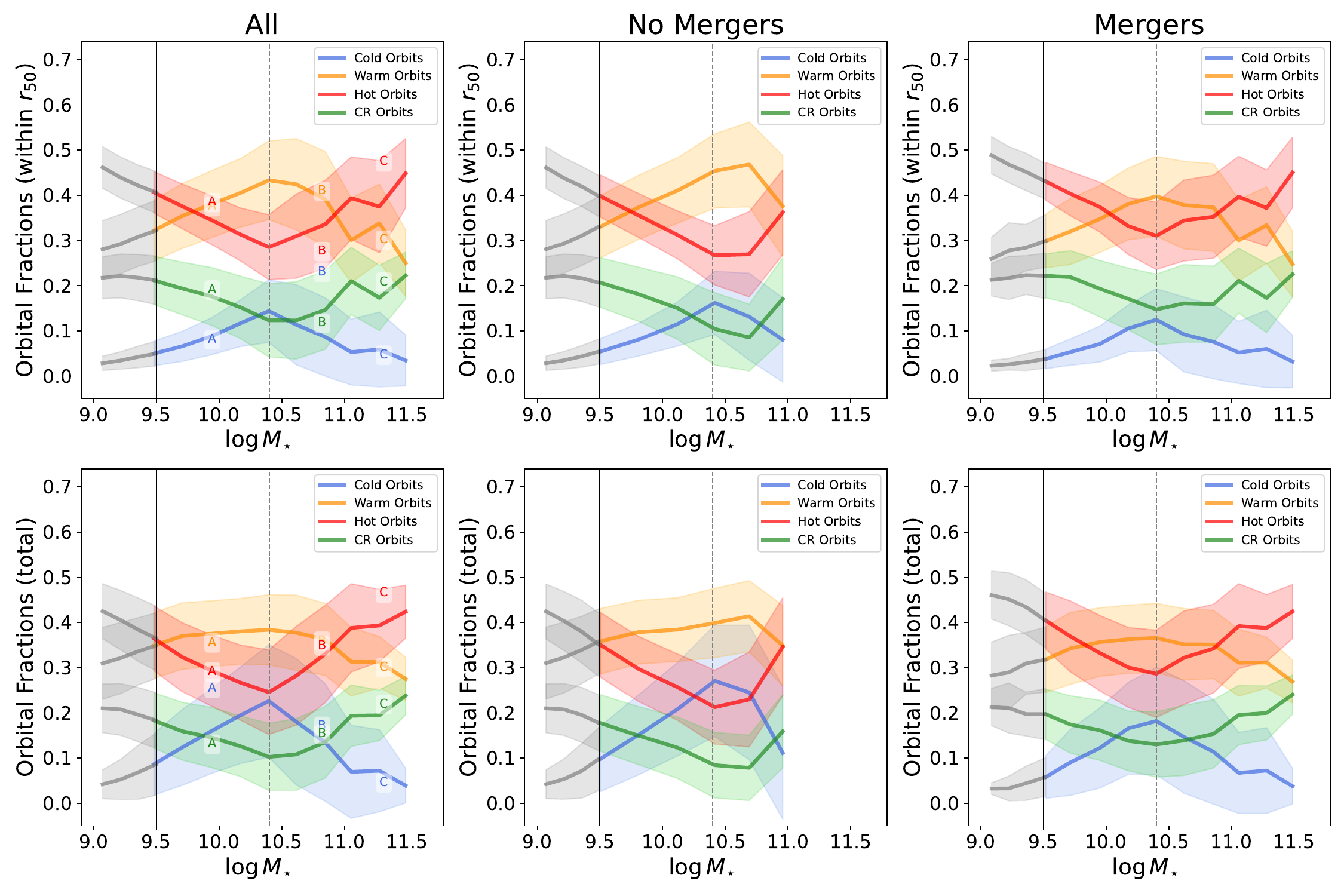}
\caption{Fraction of orbits (mass-weighted) at $z \sim 0$ as a function of stellar mass. Solid lines show the median values (for bins with at least 10 galaxies) and the shaded area delimits the 1-$\sigma$ scatter. Dashed line marks change in trends at $\log M_{\star}/\rm M_{\odot} = 10.4$. Cold orbits are shown in blue, warm orbits in orange, hot orbits in red, and counter-rotating orbits in green, as labelled. The top panels show the fraction of orbits within 1~$r_{\rm 50}$, while the bottom panels show the fractions within the whole galaxy. Left-hand panels show the results for the whole sample. The fraction of orbits for example galaxies A, B, and C (presented in Section \ref{sec:orbital_decomp}) have been labelled accordingly. We further divide our sample into galaxies that have evolved without experiencing minor or major mergers in the last 10~Gyr (between $z \sim 0$ and $z \sim 2$; middle panels), and galaxies that have undergone mergers (right-hand panels). There is a clear change in the contributions of the different types of orbits around a stellar mass of $\log M_{\star}/\rm M_{\odot} \sim 10.4$, where we see a peak in the fraction of cold and warm orbits, and a minimum in the fractions of hot and counter-rotating orbits. This change is shifted to higher stellar masses when we look at galaxies with no mergers in their evolutionary history. The total fractions of cold orbits in the whole galaxy show an overall higher contribution than those within 1~$r_{\rm 50}$.}
\label{fig:orbits_mass28}
\end{figure*}
The fractional stellar mass and $r$-band light in each orbital component are calculated within 1~$r_{\rm 50}$, by summing the mass or r-band luminosity, respectively, of each stellar particle within 1~$r_{\rm 50}$ that is in the corresponding orbital family, and dividing by the total stellar mass or $r$-band luminosity within 1~$r_{\rm 50}$. We also compute these fractions for the whole galaxy, without imposing a radial distance restriction on particles (i.e. using all the stellar particles of the subhalo).

We show the orbital distributions for three typical example galaxies - one low-mass (A), one intermediate-mass (B), and one high-mass galaxy (C) - in Fig. \ref{fig:example_1} (with stellar masses at the top of each column). These galaxies were visually selected to be representative of galaxies in their respective mass bins. For each of them, we show the binned orbital distribution of the stellar particles in the top panels. The color indicates the median stellar age of the particles in each bin.
The middle panels show the radial distribution of $\lambda_{\rm z}$ of the orbits for stellar particles in three different stellar age bins, as labelled. The bottom panels show the probability density function (PDF) of $\lambda_{\rm z}$ for particles in three different age bins. Galaxies in the low-mass range have a large fraction of their stellar mass distributed in cold and warm orbits. The cold component is also the youngest. This is typical for galaxies with stellar masses $10^{9.5} < M_{\star}/\rm M_{\odot} < 10^{10.3}$, which in {\sc Eagle} are generally supported by rotation and are actively forming stars in their disc. Intermediate-mass galaxies ($10^{10.3} < M_{\star}/\rm M_{\odot} < 10^{11}$) show a higher contribution from the hot component, but with the warm and cold components still present. These galaxies are still forming stars on average, as shown by the young population that we see concentrated in cold orbits. High-mass galaxies ($ M_{\star}/\rm M_{\odot} > 10^{11}$) are dominated by a dynamically hot and old component, extending to their outskirts. They have little contribution from cold orbits. On average, for all masses, particles in cold and warm orbits are younger than the hot component. Fractions of orbits for these galaxies can also be seen in the left-hand panels of Fig.~\ref{fig:orbits_mass28} as indicated by the labels.

\section{The stellar orbital distributions in galaxies in the local universe}\label{results}

\subsection{Stellar orbital families of {\sc Eagle} galaxies at $z=0$}

We analyse the distribution of the fraction of orbits (mass-weighted) within 1~$r_{\rm 50}$ for all galaxies at $z \sim$ 0, as a function of stellar mass (at $z \sim$ 0) in the top panels of Fig.~\ref{fig:orbits_mass28}.
\begin{figure*}
\centering
\includegraphics[scale=0.5, clip=True]{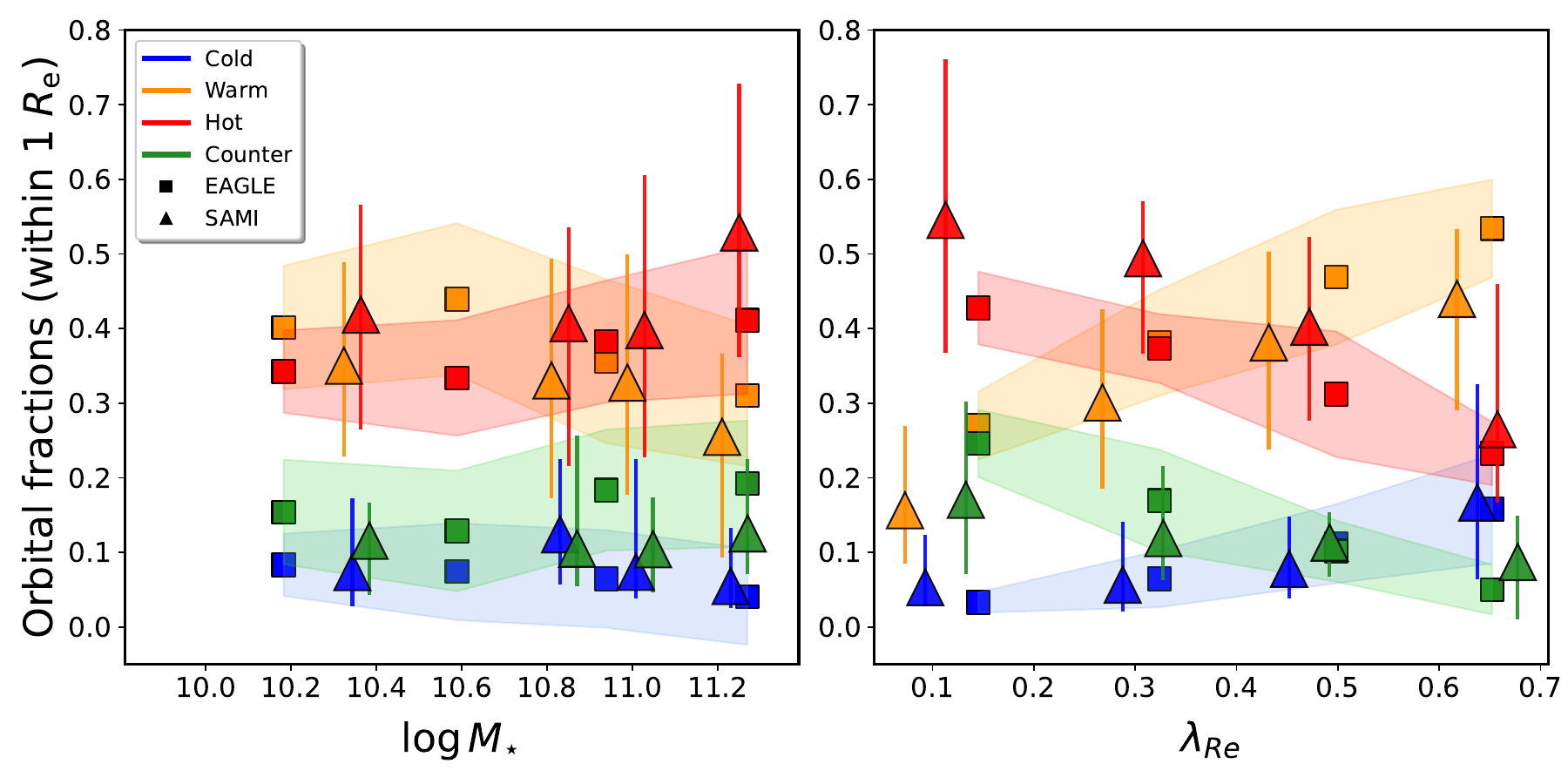}
\caption{Fraction of orbits within 1 $R_{\rm e}$ at $z \sim 0$ as a function of stellar mass (left-hand panel) and of $\lambda_{\rm Re}$ (right-hand panel). SAMI median values and scatter for each mass bin are shown as triangles with errorbars. Median values for the {\sc Eagle} stellar mass-matched passive sample are shown as squares and the shaded area delimits the 1~$\sigma$ scatter. The distributions of the fractions of orbits with stellar mass and with $\lambda_{\rm Re}$ in the {\sc Eagle} sample are in good agreement with the results from SAMI within the uncertainties.}
\label{fig:comparison_orbs}
\end{figure*}

We find that the warm and hot orbital components dominate at all masses within 1~$r_{\rm 50}$, while the median contribution of the cold and counter-rotating components is always below $\sim$ 30\%. At $\log M_{\star}/\rm M_{\odot} \lesssim 10.4$, the fraction of hot orbits decreases with increasing stellar mass, while warm and cold components become more important with increasing stellar mass, meaning that more massive galaxies are more rotationally supported than galaxies at lower masses. However, there is a clear change in the contributions of the different types of orbits at around a stellar mass of $\log M_{\star}/\rm M_{\odot} \approx 10.4$, where we see a peak in the fractions of cold and warm orbits and a minimum in the fractions of hot and counter-rotating orbits. The transition mass we find is consistent with the transition mass reported in \citet{Clauwens2018}, where the fraction of ex-situ stars starts to become more dominant in the {\sc Eagle} simulations. This transition mass is similar to what has been reported in observations to be the mass above which galaxy mergers become the dominant channel of galaxy growth in observations \citep{Kauffmann2003, Cappellari2013, Robotham2014} and {\sc Eagle} \citep{Lagos2018a}. This transition is more evident when we divide our sample into galaxies that have evolved without experiencing minor and/or major mergers in the last 10~Gyr (between $z \sim 0$ and $z \sim 2$; middle panels), and galaxies that have undergone mergers (right-hand panels). Galaxies with no recent mergers in their evolutionary history show fractions of cold and warm orbits that are steadily increasing, while the fractions of hot and counter-rotating orbits decrease with increasing stellar mass. We see evidence of a turning point at around $\log M_{\star}/\rm M_{\odot} \sim 10.75$, but the number of galaxies with no mergers at those high stellar masses is very small (112 galaxies). Even though these galaxies have not experienced any recent minor or major mergers, they have undergone very minor mergers (mass ratio $leq 0.1$). Excluding all the galaxies that also had very minor mergers removes the change in trend that we see at around $\log M_{\star}/\rm M_{\odot} \sim 10.75$. Note that for the sample of galaxies with mergers, we still see the increase of the contribution of cold orbits with stellar mass up to $\log M_{\star}/\rm M_{\odot} \sim 10.5$, but the peak contribution is smaller than what we see for galaxies that have not experienced mergers. This is expected, if we assume that mergers have an average heating effect on cold orbits (this will be explored in detail in Section~\ref{sec:discussion}). 

The trends we find when looking at the fractional contribution of different orbital families in the whole galaxies are qualitatively similar to what we find within 1~$r_{\rm 50}$. However, the total fraction of cold orbits shows an overall higher contribution than those within 1~$r_{\rm 50}$. This is not surprising since the outer regions of a galaxy are generally disc-dominated (as seen in Fig.~\ref{fig:orbits_mass28}).

In observations, stellar orbit distributions have only been derived explicitly for three large (N$>$100) samples of galaxies, in the CALIFA \citep{Zhu2018nature, Zhu2018hubble}, MaNGA \citep{Jin2020} and SAMI \citep{Santucci2022, Santucci2023} surveys, with the results being generally consistent with each other, as shown in \citep{Santucci2022}. We show the orbital fractions derived for SAMI passive galaxies, as well as the results from this work in Fig. \ref{fig:comparison_orbs}. We only select passive galaxies (with specific star formation rates $<10^{-11} \,\rm yr^{-1}$) from our {\sc Eagle} sample, stellar mass-matching it with the SAMI one. Since SAMI orbital fractions are r-band light-weighted, we derive r-band light-weighted orbital fractions for our sample of  {\sc Eagle} galaxies, to ensure a robust comparison. We show the median values of the fraction of orbits within 1~$R_{\rm e}$ (derived from the $r-$band light) for four different mass bins reported in \citet{Santucci2022} (shown as triangles, with the 1$\sigma$ scatter represented with errorbars), while median values of the light-weighted fractions in {\sc Eagle} are shown as squares, and the 1~$\sigma$ scatter in {\sc Eagle} is represented by the filled regions. The trends of the fractions of orbits that we observe with stellar mass and $\lambda_{\rm Re}$ are consistent with SAMI within the uncertainties. The inclusion of star-forming galaxies in the {\sc Eagle} mass-matched sample only has the effect of increasing the scatter in the orbits, in particular in the cold orbits (due to more prominent disc components being added to the sample). 

\begin{figure*}
\centering
\includegraphics[scale=0.45 , clip=True]{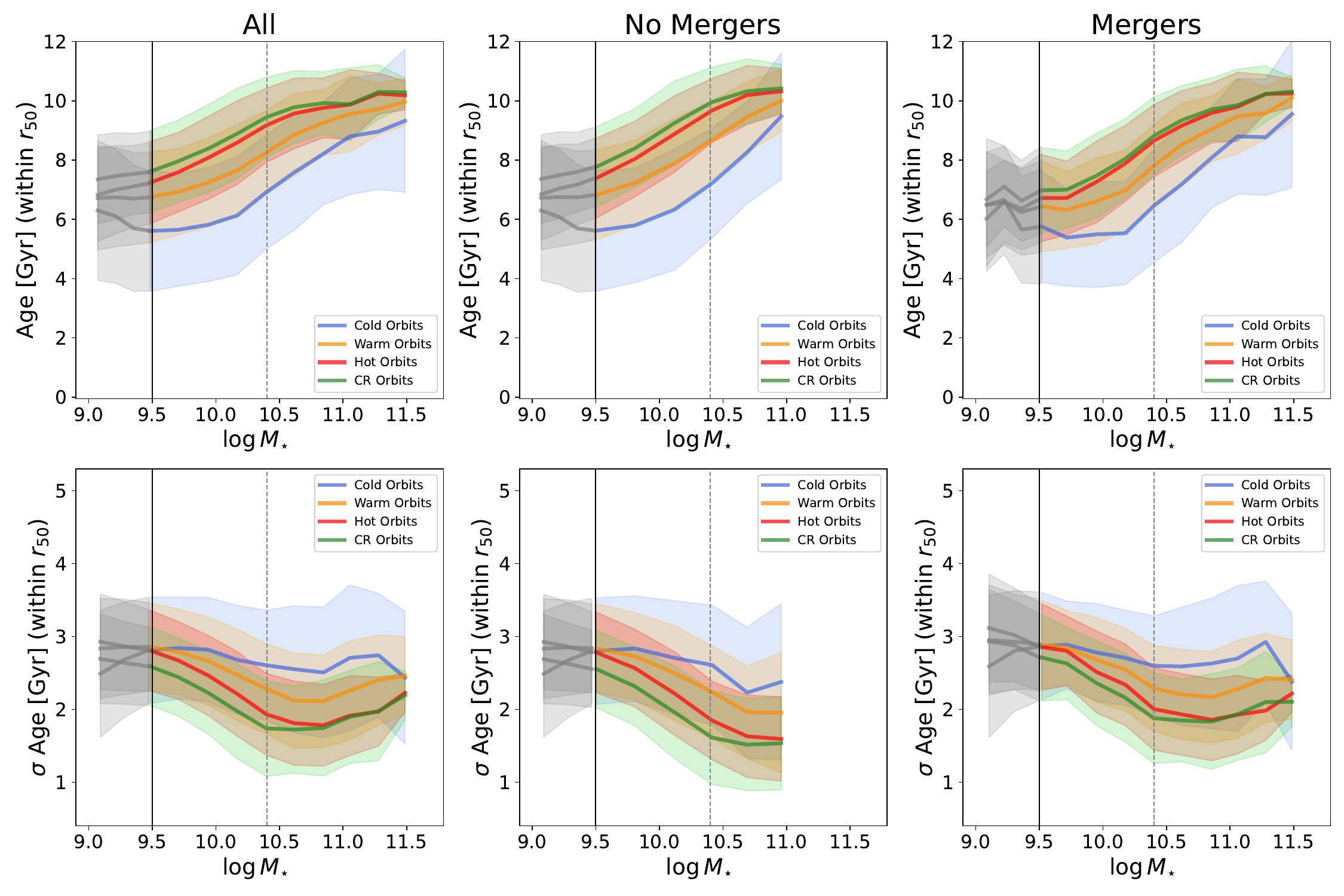}
\caption{Top panels: Median Age of stellar particles in each component within 1 $r_{\rm 50}$ at $z \sim 0$ as a function of galaxy stellar mass. Bottom panels: Median $\sigma$~Age of stellar particles in each component within 1 $r_{\rm 50}$ at $z \sim 0$ as a function of stellar mass. Lines and panels are as in Fig.~\ref{fig:orbits_mass28}. All ages increase with increasing stellar mass. Particles on cold orbits are generally younger than particles on warm orbits, with the oldest particles being on hot or counter-rotating orbits. Stellar particles on cold orbits in galaxies with no mergers are younger than those in galaxies with mergers; the opposite is seen for counter-rotating orbits, where we see older stellar particles in galaxies with no mergers compared to those with mergers. We find that  $\sigma$~Age of the cold component shows no dependence on stellar mass, while the warm, hot, and counter-rotating components have a lower $\sigma$ Age with increasing stellar mass. Particles on cold orbits have generally a wider diversity in ages than particles on warm orbits. Particles in hot and contour-rotating orbits have the lowest spread in ages.}
\label{fig:age_sigma_mass28_re}
\end{figure*}
We show the median age of the stellar particles in each orbital component in Fig.~\ref{fig:age_sigma_mass28_re}, top panels. We find that median ages of different components show qualitatively similar trends with stellar mass; stellar particles are older with increasing stellar mass regardless of their orbital family. All ages increase with increasing stellar mass, so that the most massive galaxies have the oldest populations. When looking at the different components, stellar particles on cold orbits are generally the youngest, with the oldest particles being on counter-rotating or hot orbits (which show a very similar stellar mass dependence), at all masses. However, there are some differences depending on whether galaxies had mergers or not. For example, the difference in age between the counter-rotating and hot components is very small, but slightly larger in the sample of galaxies with no merger; stellar particles on cold orbits in galaxies with no mergers are slightly younger than those in galaxies with mergers (about 1 Gyr at $\log M_{\star}/\rm M_{\odot} \sim 10.4$); the opposite is seen for counter-rotating orbits, where we see older stellar particles in galaxies with no mergers compared to those with mergers. This is not surprising since counter-rotating orbits in the sample with mergers likely result from the mergers themselves (see Section~\ref{sec:discussion} for a demonstration of that), which can happen at later times in a galaxy's life. Moreover, looking at the median ages of the total stellar particles in the galaxy (see Fig. \ref{fig:age_sigma_mass28_all}, top panels, in Appendix \ref{app:total_age}), we find that the total cold component is generally younger than the other components, in particular at low stellar masses. This difference is not as prominent when we look at the ages within 1~$r_{\rm 50}$, indicating that continuing star formation has a larger impact in the outskirts of a galaxy than within 1~$r_{\rm 50}$. Galaxies with mergers have inner cold components with slightly younger ages than galaxies with no mergers, but similar trends with mass.

The concept of ``downsizing'' in observations refers to more massive galaxies forming their stars earlier but also quicker than lower-mass galaxies \citep{Thomas2010}. With this in mind, we also analyse the standard deviation of the ages of the stellar particles ($\sigma$~Age) in each component, shown in Fig.~\ref{fig:age_sigma_mass28_re}, bottom panels. 
Even though the underlying distribution of the ages of the stellar particles is not always symmetric (discs, in particular, tend to have longer tails towards young ages), $\sigma$~Age is a simple way of comparing the spread in the ages of the stellar particles in different components. We find that $\sigma$~Age of the cold component shows no dependence on stellar mass, while the warm, hot, and counter-rotating components have lower values of $\sigma$ Age with increasing stellar mass in agreement with the ``downsizing'' scenario. The cold component almost always shows a broader range in ages compared to the warm component - indicating a more prolonged star formation history. Hot and counter-rotating components show very similar variations in ages, and display on average less spread in ages than what we see in the cold and warm orbital components at fixed stellar mass. 
We find that galaxies with no mergers (on the middle panels) show steadily decreasing $\sigma$~Age for particles in all components with increasing stellar mass. Our results do not change if we consider $\log$ Age and $\sigma \ \log$ Age instead of Age and $\sigma$ Age.

Galaxies that have experienced minor or major mergers (right-hand panels) have higher values of $\sigma$~Age for the particles in the cold component, causing the increase in $\sigma$~Age that we see in the whole sample (left-hand panel). Trends for $\sigma$~Age within 1 $r_{\rm 50}$ and for the whole galaxies are generally similar. The only noticeable difference is in the $\sigma$~Age of the cold component at low stellar masses. Within 1~$r_{\rm 50}$, the cold component has the highest values of $\sigma$~Age at all masses. However, if we consider stellar particles in the whole galaxy (see Fig. \ref{fig:age_sigma_mass28_all}, bottom panels, in Appendix \ref{app:total_age}), cold orbits have lower $\sigma$~Age than hot and warm orbits at low stellar masses. Mergers seem to lead to slightly higher $\sigma$ Age for all components.

\subsection{Stellar orbital families of {\sc Eagle} galaxies across cosmic time}
 
\begin{figure*}
\centering
\includegraphics[scale=0.33 ,trim=4cm 6cm 4cm 7cm, clip=True]{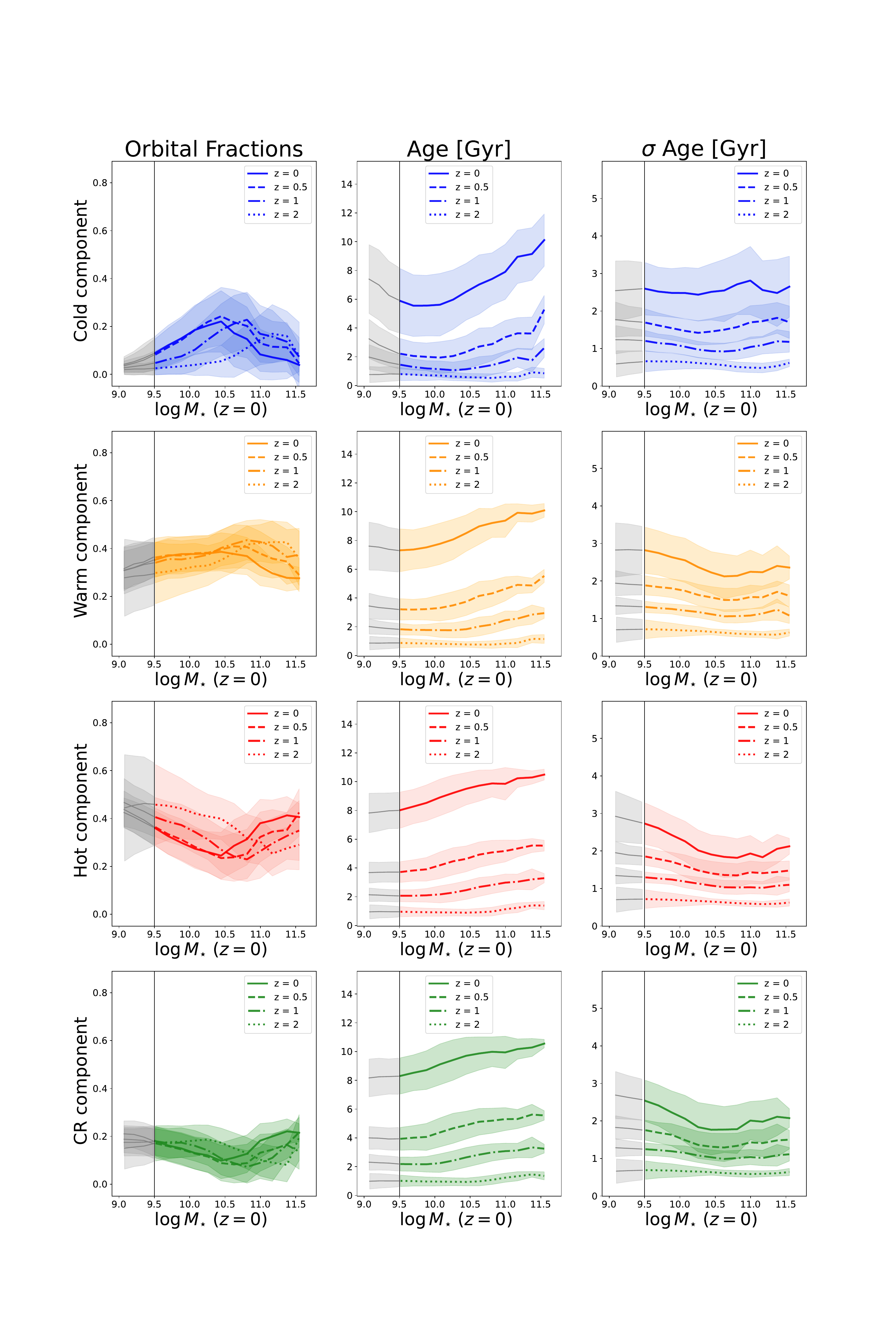}
\caption{Orbital fractions (panels on the left), stellar Age (middle panels), and $\sigma$~Age (right-hand panels) as a function of the $z \sim 0$ galaxy stellar mass at different redshifts, as labelled in each panel. Cold orbits are shown in blue (top panels), warm orbits in orange (second from top panels), hot orbits in red (second from bottom panels), and counter-rotating orbits in green (bottom panels). We show the median values for each component at different redshift: $z \sim 0$ as a solid line, $z \sim 0.5$ as a dashed line, $z \sim 1$ as a dash-dotted line, and $z \sim 2$ as a dotted line. Shaded regions indicate the 1~$\sigma$ scatter from the median. Galaxies with $M_{\star}/\rm M_{\odot}>10^{10.5}$ at $z=0$, have a peak in the fraction of cold and warm orbits at $z \sim 1$, with the most massive galaxies reaching their peak earlier ($z \sim 2$), and then steadily decrease with time. Opposite trends with time are found for hot and counter-rotating orbits, so that the orbital fractions are higher at $z=0$ for high-mass galaxies. For galaxies with $M_{\star}/\rm M_{\odot}<10^{10.5}$ at $z=0$, the fraction of cold and warm orbits steadily increases from $z=2$ to $z=0$, albeit with little evolution between $z=0$ and $z=0.5$. Hot and counter-rotating orbits, on the other hand, decrease with decreasing redshift. The ages of the components increase with increasing redshift. $\sigma$ Age decreases going towards high masses for all but the cold orbits, where it stays the same.}
\label{fig:orbits_redshift}
\end{figure*}

We now explore how the orbital structures of galaxies change with redshift. We trace the main progenitors of each $z=0$ galaxy in our sample, and we calculate the fraction of their orbital components at $z=0.5$, $z=1$, and $z=2$ the same way as we did for our $z=0$ sample. We show the total fraction of orbits in each component as a function of the stellar mass that the galaxies have at $z = 0$ for different redshifts in Fig. \ref{fig:orbits_redshift}, left-hand panels. Note though, that the fractions themselves are calculated relative to the stellar mass of the progenitor galaxy at the respective redshift. We only show the total fractions of orbits since the behaviour within 1~$r_{\rm 50}$ is very similar to that in the whole galaxy. The idea of showing the fraction of orbits as a function of the stellar mass at $z=0$ is to isolate the evolution in the orbital fractions that happens in the $y$-axes. We note that, if we consider all the galaxies in our sample, regardless of whether they survived to $z=0$, we find similar trends, albeit with slightly weaker overall evolution. Similarly, if we split the sample between galaxies that by $z=0$ had mergers or not, we see overall qualitatively the same differences discussed in the context of the $z=0$ distributions (Fig. \ref{fig:orbits_mass28} and \ref{fig:age_sigma_mass28_re}), and hence here we limit to showing the full sample without splitting by merger history.

Galaxies with $M_{\star}/\rm M_{\odot}>10^{10.5}$ at $z=0$, have a peak in the fraction of cold and warm orbits at $z \sim 1$, with the most massive galaxies reaching their peak earlier ($z \sim 2$), and then steadily decrease with time, presumably due to the merging activity in these galaxies which leads to an increase in the fraction of hot and counter-rotating orbits.
In the case of hot and counter-rotating orbits, we obtain the opposite trends with time, where the orbital fractions are higher at $z=0$ for high-mass galaxies. Most of the evolution in the fractions of orbits (cold, warm, hot, and counter-rotating) in massive galaxies ($\gtrsim 10^{10.5} M_{\odot}$, happens between $z=0.5$ and $z=0$, presumably due to their active merger history. 
The MUSE MAGPI survey \citep{Foster2021} is the ideal survey to test these predictions, since it focuses on studying galaxy transformations between $z=0$ and $z \sim 0.4$ in massive galaxies.

For galaxies with $M_{\star}/\rm M_{\odot}<10^{10.5}$ at $z=0$, we see that their fraction of cold orbits steadily increases from $z=2$ to $z=0$, albeit with little evolution between $z=0$ and $z=0.5$. This is reminiscent of the evolutionary pathways identified by \cite{Lagos2017} for galaxies that at $z=0$ have high specific angular momentum in {\sc Eagle}: most of their spin-up happens at $z \lesssim 1$. Similar behaviours are also found for the fractions of warm orbits, albeit with an overall weaker evolution at $M_{\star}/\rm M_{\odot}<10^{10.5}$ than what we see for cold orbits. Hot and counter-rotating orbits, on the other hand, decrease with decreasing redshift. For these galaxies (with $M_{\star}/\rm M_{\odot}<10^{10.5}$ at $z=0$) we see that most of the increase in their orbital fractions happens between $z=2$ and $z=1$, with little evolution at $z<1$. In fact, these galaxies display almost no evolution in their orbital fractions from $z=0.5$ to $z=0$. 

These results suggest that at $z \sim 2$ only the progenitors of very massive $z=0$ galaxies ($M_{\star}\gtrsim 10^{11}\,\rm M_{\odot}$) had their disc components in place. For galaxies with $10^{10.3}\lesssim M_{\star}/\rm M_{\odot}\lesssim 10^{11}$, the disc primarily forms between $z=2$ and $z=1$, while for galaxies of stellar mass $<10^{10}\,\rm M_{\odot}$, discs are formed preferentially at $z<1$. 

We show the median age of the stellar particles in each component as a function of stellar mass at $z=0$, $z=0.5$, $z=1$, and $z=2$ in the middle panels of Fig.~\ref{fig:orbits_redshift} and ~$\sigma$ Age for each component in the right-hand panels. At fixed redshift, the ages of the components increase with increasing stellar mass. The degree of increase with stellar mass changes with redshift, becoming steeper at lower redshifts. Similarly, at fixed stellar mass, the ages increase with decreasing redshift. The trends still hold if we normalise the ages of each component by the age of the Universe at each redshift (as seen in Appendix \ref{app:hubbletime}, Fig. \ref{fig:htime}). Cold orbits are always the youngest component in galaxies at all redshifts, although very close to the ages of the other components at $z=2$. We also note that the counter-rotating component is the oldest at all redshifts, closely followed by the hot component. The warm component is always younger than the hot component, but older than the cold one. This is consistent with observations that found evidence of older stars being on hotter orbits than young stars in present-day systems such as the Milky Way (e.g. \citealt{Bird2013}), M31 (e.g. \citealt{Quirk2018}), and in MaNGA (\citealt{Shetty2020}) and SAMI \citep{Foster2023} galaxies.
Interestingly, in our sample, this is true not only at all redshifts, but also at all stellar masses (as seen in the top panels of Fig.~\ref{fig:age_sigma_mass28_re}). This indicates that the formation of stars in the different orbital components happens in similar ways for all galaxies: the counter-rotating and hot components are the first to cease star formation while the cold component is the last to form. 
Whether this means that the components are already dynamically in place, since stellar particles can be added (or migrate) to different components due to mergers and galaxy interactions, is still an open question. We also see a decrease in the scatter of the ages going towards high stellar masses for all components but the cold orbits, where we see a fairly stellar-mass independent scatter in stellar ages. This is consistent with stars in more massive galaxies forming earlier and over a shorter timescale (i.e. ``downsizing'';  \citealt{Thomas2010}).
For $z \gtrsim 1$, $\sigma$~Age does not show any trend with stellar mass, while we see a decrease in the $\sigma$~Age with increasing stellar mass for the warm, hot, and counter-rotating components at $z=0$. Cold orbits do not show any decrease in $\sigma$~Age with stellar mass, even at $z=0$. 

\section{Discussion}\label{sec:discussion}
We have analysed the orbital distributions of galaxies at $z \sim 0$, with stellar masses above $\log M_{\star}/\rm M_{\odot} \sim 9.5$ in the {\sc Eagle} simulations. We divided each galaxy into four different orbital components, depending on their orbital circularity: a cold component (with close to circular orbits), a warm component, a hot component, and a counter-rotating component. We calculated the fraction of stellar mass and the median stellar age in each component. We find that changes in the orbital distributions are mostly correlated with the stellar mass of the galaxies, as previously seen in the literature (e.g., \citealt{Zhu2018nature, Jin2020, Santucci2022}). We also find differences in the ages and orbital fractions depending on the galaxy merger history. 
\begin{figure*}
\centering
\includegraphics[scale=0.35 , clip=True]{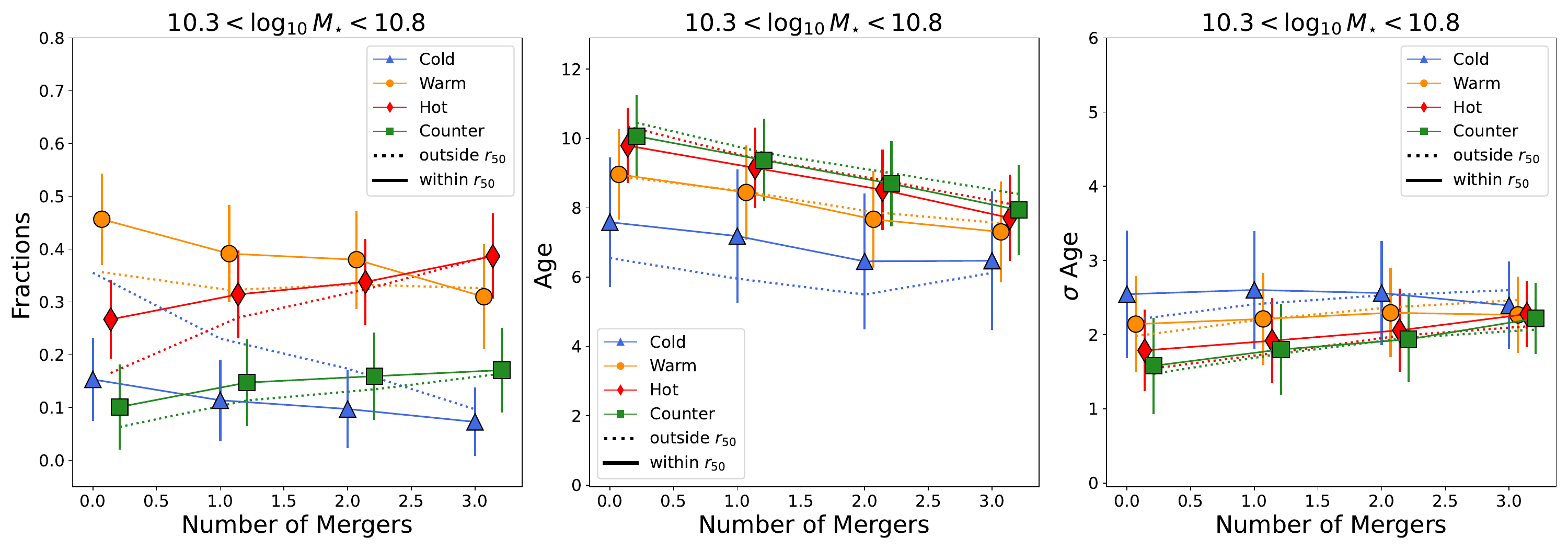}
\caption{{\it Left-hand panel:} Medians (symbols connected by lines) of the stellar mass fractions in each orbital component, as labelled, within 1~$r_{\rm 50}$ (solid lines) and outside 1~$r_{\rm 50}$ (dotted lines) for galaxies $ 10.3 < \log M_{\star}/ \rm M_{\odot} < 10.8$ at $z=0$, as a function of the total number of minor plus major mergers experienced by the galaxies. The latter is the cumulative number of merger events at $0\le z\le 2$.  
Errorbars show the 1-$\sigma$ scatter around the median. 
Note that we perturb the x-axis position of the symbols slightly to avoid overlapping errorbars.
The stellar mass fraction in warm and cold orbits decreases with increasing number of mergers, while the fractions of hot and counter-rotating orbits increase with increasing number of mergers. 
{\it Middle panel:} As in the left-hand panel but for the mean stellar age of each orbital component. 
Particles in galaxies that have undergone a higher number of mergers have younger ages, independently of the component they belong to, with the cold component having the youngest ages overall. 
{\it Right-hand panel:} As is the left-hand panel but for 
%
$\sigma$~Age of each orbital component. There is an increase in $\sigma$~Age for the stellar particles in each orbital component with increasing numbers of mergers, except for cold orbits, where $\sigma$~Age remains approximately constant.}
\label{fig:all_pars_nmergers_mid}
\end{figure*}
\subsection{Number of wet and dry mergers}
In order to constrain the correlation between galaxy mergers and a galaxy's orbital properties, we show in the left-hand panel of Fig.~\ref{fig:all_pars_nmergers_mid} the median values of the orbital fractions within (solid lines) and outside 1~$r_{\rm 50}$ (dotted lines) as a function of the number of minor and/or major mergers each galaxy has undergone since $z=2$, for galaxies of intermediate stellar masses ($ 10.3 < \log M_{\star}/ \rm M_{\odot} < 10.8$). Lower and higher mass galaxies show similar global trends, but with different absolute values (see Fig. \ref{fig:all_pars_nmergers_low} and Fig. \ref{fig:all_pars_nmergers_high} in Appendix \ref{app:mergers}). 
We see a decrease in the fractions of warm and cold orbits with increasing number of mergers, while the fractions of hot and counter-rotating orbits increase with increasing number of mergers. The decrease of cold orbits with the number of mergers is stronger when we consider the region outside $1\,r_{\rm 50}$, instead of only what is enclosed within $1\,r_{\rm 50}$. The conjecture is that mergers tend to disrupt cold and warm orbits, which results in an increase in the fraction of hot and counter-rotating orbits after a galaxy has experienced a merger. The effect is cumulative, as 
galaxies with the highest fractions of counter-rotating and hot orbits are those with a cumulative number of mergers $\ge 3$.
This result is consistent with a two-phase formation scenario \citep{White1980,Kobayashi2004, Oser2010}, where more massive galaxies (which are also the galaxies showing high fractions of counter-rotating and hot orbits), start forming stars at earlier epochs than less massive ETGs \citep{DeLucia2006, Dekel2009}. After star formation is quenched, a second phase of slower evolution is dominated by mergers, leading to the increase of the hot and counter-rotating fractions. 
This is also what was shown to happen in {\sc Eagle} for slowly rotating galaxies by \citet{Lagos2022}. They show that the vast majority of slow-rotating galaxies experience quenching before they suffer the major morphological transformation that leads to their slowly-rotating nature.

Ages of stellar particles as a function of the number of major and/or minor mergers are shown in the middle panel of Fig. \ref{fig:all_pars_nmergers_mid}. Particles in galaxies that have undergone a higher number of mergers have younger ages, independently of the component they belong to, with the cold component having the youngest ages overall. This is likely a consequence of any gas present being funneled to the galaxy centre and leading to new episodes of star formation by the galaxy merger.
In addition, accreted stars from the secondary galaxy are likely to be younger than those in the primary due to the underlying positive correlation between stellar age and mass. This is also visible by looking at the age dispersion (Fig. \ref{fig:all_pars_nmergers_mid}, right-hand panel), where we see an increase in $\sigma$~Age for the stellar particles in each component, but not for cold orbits, with increasing number of mergers. 

Different types of mergers during the second phase can have different effects on a galaxy's structure and composition, leading to the formation of slow- or fast-rotating galaxies. \citet{Cappellari2016} described how these galaxies can form through two main formation channels: fast-rotating galaxies start as star-forming discs and evolve through a set of processes dominated by gas accretion, bulge growth, and quenching. By comparison, slowly rotating galaxies assemble near the centre of massive haloes via intense star formation at high redshift \citep{Naab2009,Clauwens2018}, and evolve from a set of processes dominated by gas-poor mergers, resulting in more triaxial shapes. Gas-poor and gas-rich mergers can also have very different effects on the spin of galaxies (as shown by \citealt{Naab2014} and \citealt{Lagos2022}, for example). 

To explore how orbital fractions are influenced by different types of mergers, we classify the mergers into gas-rich (wet) and gas-poor (dry) using the ratio of gas to stellar mass of the merger ($f_{\rm gas,merger}$, derived by \citealt{Lagos2018a}). The latter is the sum of the gas masses of the two merging galaxies divided by the sum of the stellar masses of the two galaxies. In this paper, we classify mergers with a gas fraction below $0.2$ as dry or gas-poor, and mergers with gas fractions above this value as wet or gas-rich. We note that changing the threshold does not impact the results, since the distribution of the gas fraction is highly bimodal: dry mergers have a median gas fraction of $\sim 0.10$, while wet mergers have a median gas fraction of $\sim 0.98$. 

\begin{figure}
\centering
\includegraphics[scale=0.45 , trim=0.5cm 3cm 2.0cm 3.5cm,clip=True]{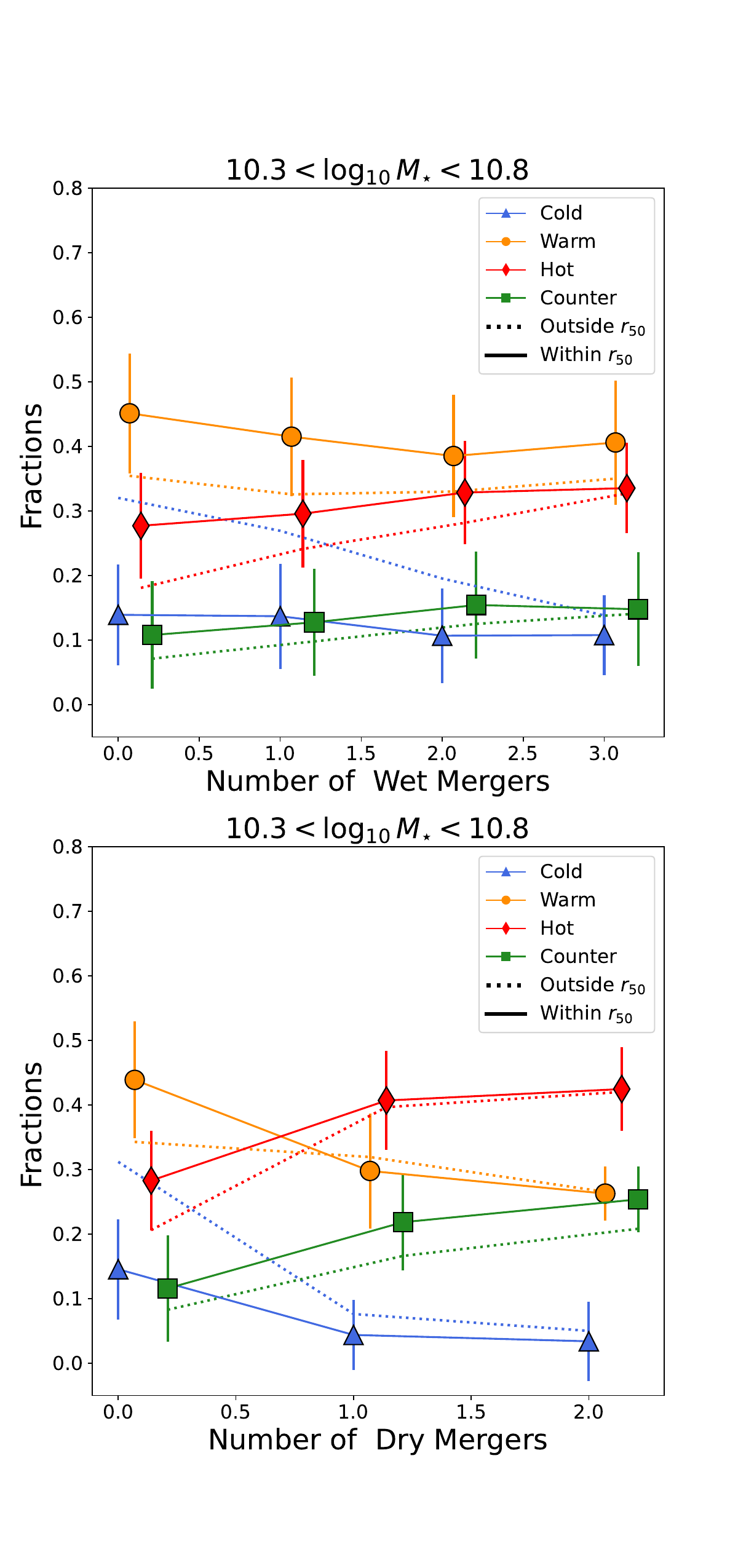}
\caption{As in the left-hand panel of Fig.~\ref{fig:all_pars_nmergers_mid}, but as a function of only the number of wet (top panel) or dry (bottom panel) minor or major mergers experienced by a galaxy. 
Dry mergers have a significantly greater impact on orbital fractions than wet mergers. Hot and counter-rotating orbits show an increase of 50-100\% after only one dry merger compared to 10-20\% for one wet merger. The effects of wet mergers on the orbital fractions are weaker, but appear to accumulate with the number of wet mergers.}
\label{fig:orbs_wetdry}
\end{figure}

We show the median values of the fractions of orbits as a function of the number of wet and dry mergers in Fig. \ref{fig:orbs_wetdry}. Note that galaxies that experienced both wet and dry mergers appear in both panels; selecting galaxies with dry mergers only or wet mergers only does not change the trends we see here. In general, the fractions of orbits follow the same qualitative trends we see in Fig.~\ref{fig:all_pars_nmergers_mid}. However, the effect of wet mergers is weak and leads to a small change in the orbital fractions of up to 20\%. 
Their effect, however, appears to accumulate with the number of wet mergers. Galaxies having undergone $\ge 3$ wet mergers have $\sim$ 15 percent more stellar mass in hot and counter-rotating orbits than galaxies with $=1$ wet merger. 

Dry mergers, on the other hand, seem to have a great impact on the amount fractions change. In particular, one dry merger is enough to see a large increase (50-100\%) in the fractions of hot and counter-rotating orbits (and a decrease in the fractions of warm and cold orbits). This is not only true within 1~$r_{\rm 50}$, but also in the outer regions, where the effects are stronger. Interestingly, we see that in general galaxies have fractions of hot orbits above $0.4$ and counter-rotating orbits above $0.2$ only if they have experienced at least one dry merger in their evolution (since $z=2$). Similarly, dry mergers seem to be the main channel to reduce the contributions from warm and cold orbits (with the fractions of warm orbits below $0.3$ and cold orbits below $0.1$). We do not show the effects of dry and wet mergers on the ages of the components here, since we do not see significant differences for the different components (we find slightly younger ages, for all components, with increasing number of wet mergers, and slightly older ages for the cold component with increasing number of dry mergers). We do not find significant differences when we look at the fractions of orbits as a function of minor and major dry mergers.

\subsection{Ex-situ fractions}\label{sec:orgins_orbits}

The internal orbital structures of the galaxies in our sample are also consistent with a scenario where a massive galaxy's evolution is dominated by galaxy mergers, leading to their average spin-down. Previous results from simulations suggest that stars on different orbits have different formation paths. The cold components are mostly young stars formed in-situ, the warm component likely traces old stars formed in-situ or stars being heated from cold discs via secular evolution, and a small fraction of the warm component stars could be accreted \citep{Gomez2017,Park2021}. The stars on hot orbits in the outer regions should mostly be accreted via minor or major mergers \citep{Gomez2017, Tissera2017}, while stars on hot orbits in the inner regions are predicted to have formed in-situ at high-redshift \citep{Williams2014,Du2021,Yu2021}. However, the origin of the hot orbits, particularly in galaxies that have not experienced mergers, are predicted to be different for low- and high-mass galaxies: in low-mass galaxies, stellar particles are mostly born hot; while in high-mass galaxies, they are born in cold/warm orbits, but heated by bar-like secular evolution. For example, \cite{Yu2023Fire} found that particles on hot orbits in Milky Way-like galaxies in the FIRE-2 cosmological zoom-in simulations \citep{Hopkins2018Fire} are all born hot. This is an interesting aspect that requires further investigation. However, the simple assumptions made in {\sc Eagle} regarding the structure of the interstellar medium, specifically the fact that gas does not cool down below 8,000~K, and therefore by construction is bound to have a significant dispersion component in galaxies of stellar masses $\lesssim 10^{9.5}\,\rm M_{\odot}$, together with resolution limitations, prevent us from studying how orbits could transition from cold to warm in low-mass galaxies.


We find that massive galaxies have a substantial amount of stellar mass on hot orbits (over $40$\% and increasing with increasing stellar mass) in the inner regions (within $1\,r_{\rm 50}$), with stellar ages \(>\)10 Gyr. We also have high fractions of hot orbits when considering the whole galaxy. We also find that the most massive galaxies have smaller ranges in the ages of their particles (i.e. stars formed at around the same time) in the inner regions, and that these galaxies, being less supported by rotation, also generally have older mean ages than less massive (more rotationally-supported) galaxies. 

\begin{figure*}
\centering
\includegraphics[scale=0.4, clip=True]{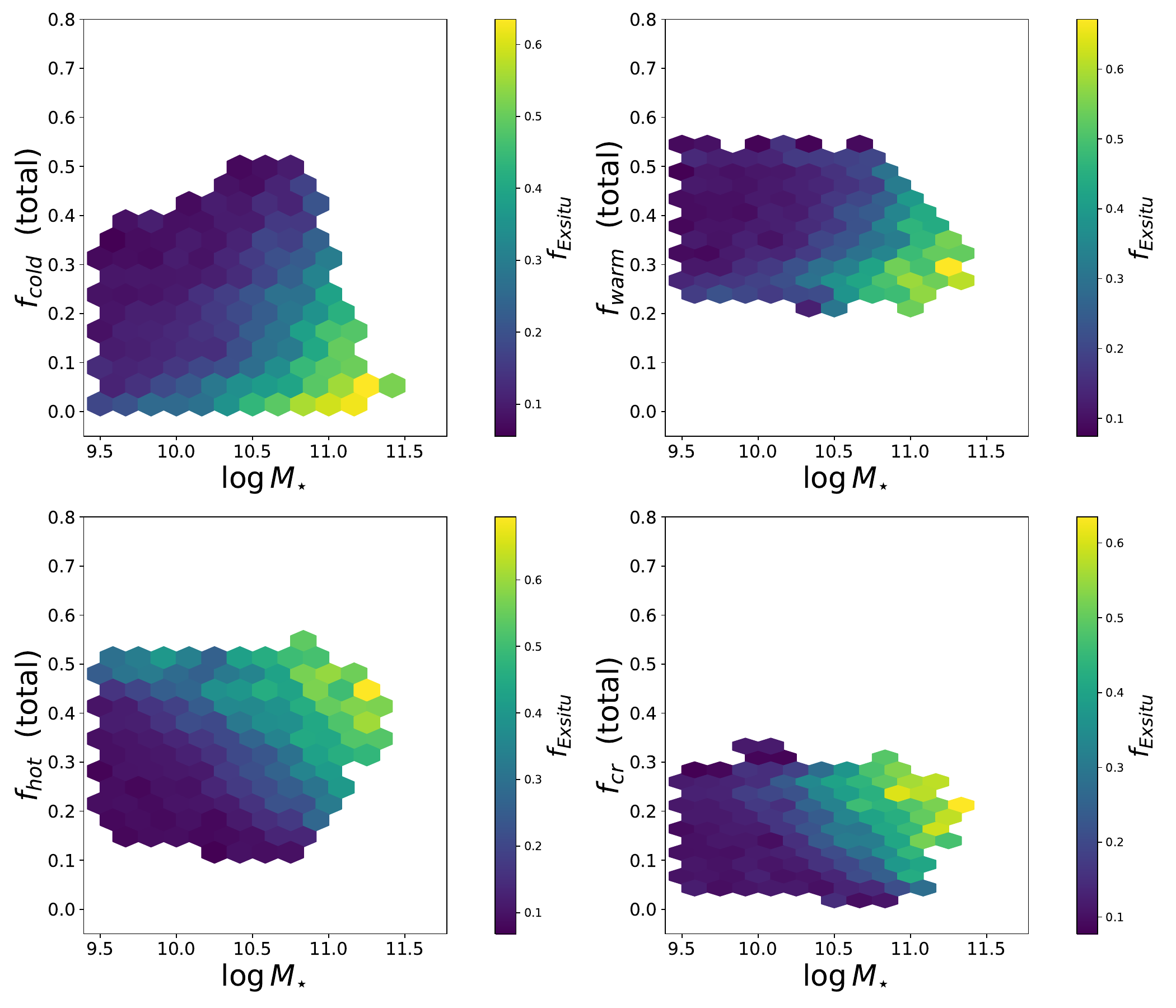}
\caption{Fraction of orbits in different components in the whole galaxy as a function of stellar mass for galaxies at $z=0$, colour-coded by the median ex-situ fraction of the galaxies in each 2D bin, as labelled in the colour bar. A minimum of 5 galaxies is required for each bin. 
Low-mass galaxies ($ \log M_{\star}/ \rm M_{\odot} <$ 10.5) have small
ex-situ fractions, except for the low-mass galaxies with very high fractions of hot orbits ($\gtrsim 0.5$). 
The ex-situ fractions display
a sharp increase around $ \log M_{\star}/ \rm M_{\odot} \sim 10.5$. For stellar masses above $ \log M_{\star}/ \rm M_{\odot} \sim 10.6$, at fixed stellar mass, galaxies with higher ex-situ fractions have higher fractions of hot and counter-rotating orbits and lower fractions of cold and warm orbits.}
\label{fig:exsitu}
\end{figure*}

To better constrain the correlation between orbital components and ex-situ fractions, we show in Fig.~\ref{fig:exsitu} the fraction of orbits in different components in the whole galaxy as a function of stellar mass, divided into hexabins colour-coded by the median of the total ex-situ fractions (derived by \citealt{Davison2020}) for the galaxies included in the bin. A minimum of 5 galaxies is required for each bin. Low-mass galaxies ($ \log M_{\star}/ \rm M_{\odot} <$ 10.5) have very 
small 
ex-situ fractions (with typical values being less than 0.1), except for galaxies with hot orbital fractions $\gtrsim 0.5$ where the ex-situ fraction increases to $\approx 0.36$.
The ex-situ fractions have a sharp increase around  $ \log M_{\star}/ \rm M_{\odot} \sim 10.5$, as previously seen in {\sc Eagle} (e.g. \citealt{Clauwens2018}). Fractions of orbits within 1~$r{\rm 50}$ show similar, but weaker, trends and are shown in Appendix \ref{appendix:exsitu_re}. This is not surprising since Fig. \ref{fig:all_pars_nmergers_mid} shows that mergers and galaxy interactions have a stronger effect on the outskirts of galaxies (also seen by \citealt{Lagos2018a,Karademir2019}).

We find that low mass ($ \log M_{\star}/ \rm M_{\odot} <$ 10.5) in-situ dominated galaxies with negligible ex-situ fractions are characterised by high fractions of cold and, in particular, warm orbits. In galaxies with larger masses ($\log M_{\star}/ \rm M_{\odot} >$ 10.5) the galaxy structure is altered by the accreted stars and gas from merger events in a way that is strongly dependent on stellar mass. It is clear, however, that there are correlations between the ex-situ fraction and the fraction of mass in different orbital families at fixed stellar mass 
for galaxies with stellar masses above $ \log M_{\star}/\rm \rm M_{\odot} \sim 10.6$. In these galaxies, higher ex-situ fractions are associated with lower fractions of cold and warm orbits and higher fractions of hot and counter-rotating orbits. Since the fraction of orbits and the stellar mass of a galaxy are measurable quantities (using advanced modelling techniques), we can use them to help us to put constraints on the galaxy's ex-situ fraction.

Our results suggest that galaxies with $ \log M_{\star}/ \rm M_{\odot} < 10.5$ in {\sc Eagle} are rotationally supported, dominated by young ($\sim$ 6 Gyr) stellar particles in cold and warm orbits, which are driven by in-situ star formation ($>$90\% of the stellar particles in these galaxies are formed in-situ). We also find a small fraction of galaxies in this stellar mass range where stellar particles formed ex-situ contribute to enhance the fraction of hot-orbits (as shown in Fig.~\ref{fig:all_pars_nmergers_mid}). 

Above $\log M_{\star}/\rm  \rm M_{\odot} \sim 10.5$ we see a change in the contributions of the different components (Fig.~\ref{fig:orbits_mass28}), so that galaxies transform towards a more pressure-supported configuration (dominated by old stellar particles in hot orbits). This transformation seems to be generally driven by dry mergers (Fig.~\ref{fig:orbs_wetdry}), which is associated with accretion from stars formed ex-situ (Fig.~\ref{fig:exsitu}).

\subsection{Counter-rotating orbits}
The establishment of counter-rotating orbits, which are generally composed of old stellar particles, is thought to happen at a much earlier stage of galaxy evolution, during which both prograde and retrograde mergers contribute nearly equally to galaxy bulge growth. However, the total fraction of counter-rotating orbits should decrease with redshift, when the galaxy halo effectively accretes ambient cold gas, which cools and eventually forms a coherent gaseous and stellar disc (see e.g., \citealt{Driver2013, Pillepich2019}). For high-mass galaxies ($ \log M_{\star}/ \rm M_{\odot} >$ 10.5), we see an increase in the fractions of counter-rotating orbits (Fig. \ref{fig:orbits_redshift}), generally due to dry mergers (as seen in Fig.~\ref{fig:orbs_wetdry}).

\begin{figure*}
\centering
\includegraphics[scale=0.4 , clip=True]{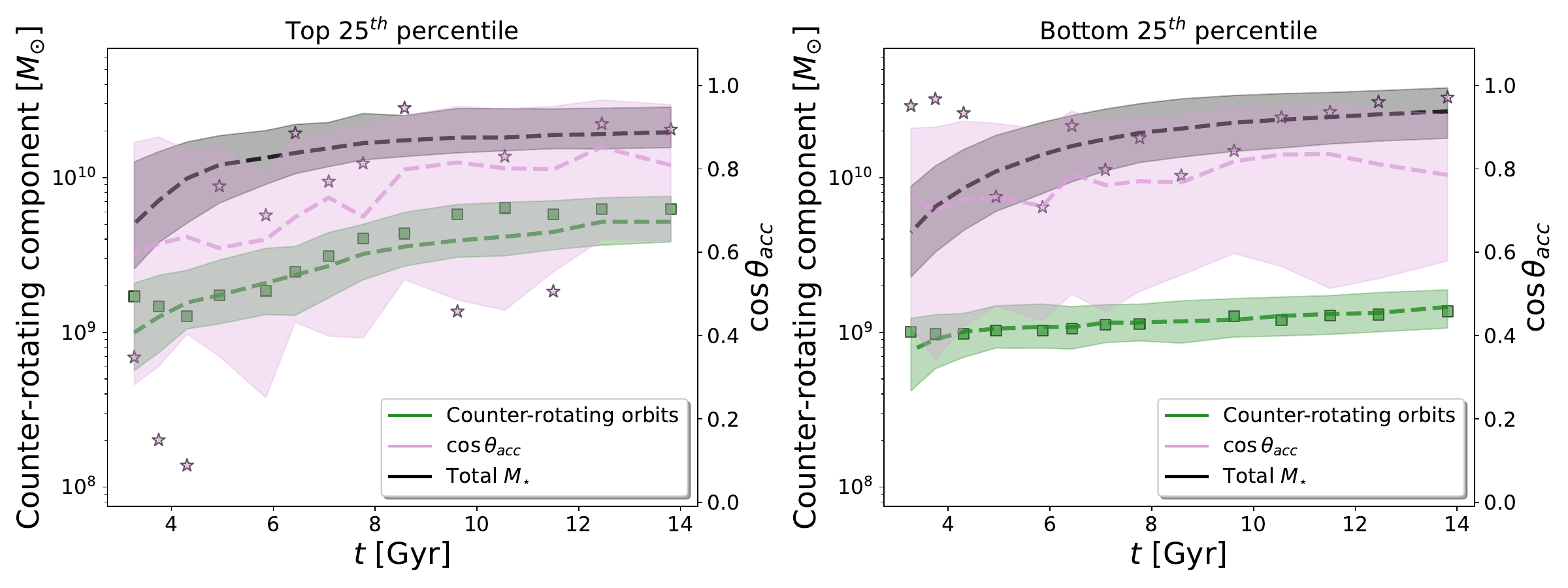}
\caption{{\it Left-hand panel:} stellar mass of the counter-rotating component (in green), $\cos$ of the inflow angle (in plum), and total stellar mass (in black) as a function of time (in Gyr) for central galaxies with no mergers in the top 25$^{\rm th}$ percentile of the distribution of counter-rotating fractions for galaxies at $z \sim 0$ with stellar masses above $\log M_{\star}/ \rm M_{\odot} \sim 10.1$. Dashed lines show the median values at each redshift and the shaded area delimits the 25$^{\rm th}$ and 75$^{\rm th}$ percentile ranges. Green squares represent the mass of the counter-rotating component for one example galaxy in each population and plum stars represent the inflow accretion angle ($\theta_{acc}$). We show these example galaxies to provide an example of the variation of the alignment of the gas inflow and the mass in the counter-rotating component with time in individual galaxies. A value of $\theta_{acc}<20^{\circ}$ means that the inflow of gas is closely aligned with the rotation axis of the galaxy.
For these galaxies, see an increase in the stellar mass of the whole galaxy, as well as in the stellar mass in the counter-rotating component. 
{\it Right-hand panel:} lines and colours as the left-hand panel, but for central galaxies with no mergers in the bottom 25$^{\rm th}$ percentile of the distribution of counter-rotating fractions for galaxies at $z \sim 0$ with stellar masses above $\log M_{\star}/ \rm M_{\odot} \sim 10.1$. For these galaxies, we see an increase in the stellar mass of the whole galaxy, similar to the increase we see for the population in the top 25$^{\rm th}$ percentile. However, we do not see any significant increase in the stellar mass of the counter-rotating component. 
Inflows with inclination angles highly misaligned ($\cos \theta_{acc}<$0.8) from the rotation axis of the disc at high redshift seem to support the formation of stellar particles in counter-rotating orbits. }
\label{fig:fcr_angle}
\end{figure*}

However, we have cases ($\sim$ 8\%) where galaxies with no minor or major mergers in the last 10 Gyr (since $z \sim 2$) have fractions of counter-rotating orbits comparable to those of galaxies that have experienced mergers, pointing to the need for an alternative path for the formation of counter-rotating orbits. Previous numerical studies \citep[e.g.,][]{Scannapieco2009,Clauwens2018,Garrison-Kimmel2018} and cosmological simulations \citep[e.g.,][]{Park2019} found that galaxies sometimes develop counter-rotating discs due to the gas infalling in a direction misaligned
with the existing co-rotating disc plane. Therefore we suspect a possible way to form these orbits is through inflows of cold gas at particular angles. To test our hypothesis, we select a sample of central galaxies at $z\sim0$ with stellar mass above $\log M_{\star}/ \rm M_{\odot} \sim 10.1$. We then calculate the median value of the fraction of counter-rotating orbits for these galaxies and select galaxies that are in the top (highest fractions of counter-rotation) and bottom 25$^{\rm th}$ (lowest fractions of counter-rotation) percentiles. We exclude satellite galaxies since their tidal interactions may contribute to the counter-rotating orbits. The mass threshold is selected in order to have kinematic quantities robustly measured. Selecting a different mass limit - for example $\log M_{\star}/ \rm M_{\odot} \sim 10.6$ - does not change our conclusions. From these two populations (top and bottom 25$^{\rm th}$ percentiles), we exclude all the galaxies that have not experienced minor or major mergers since $z=2$. For the remaining galaxies in these two subsamples we show the median evolution of the mass of their counter-rotating as a function of time (in Gyr), as well as their total stellar mass and the cosine of the inclination angle of the accreted cold gas (as derived by \citealt{Jimenez2022}), in Fig.~\ref{fig:fcr_angle}. To provide an example of the variation of the alignment of the gas inflow and the mass in the counter-rotating component with time in individual galaxies, we show values for an example galaxy in each population (green squares for the fraction of counter-rotating orbits and plum stars for the $\cos \theta_{acc}$). A value of $\cos \theta_{acc}$ close to 1 means that the inflow of gas is closely aligned with the rotation axis of the galaxy.  We find that the misalignment between the inflowing gas and the disc’s rotation axis, especially in the early Universe, can have a strong impact on the formation of counter-rotating orbits. Both populations have, in general, similar net accretion rates since their median total stellar mass distributions show a similar increase with time (black dotted lines in Fig.~\ref{fig:fcr_angle}). Inflows with inclination angles highly misaligned ($\cos \theta_{acc} <$0.8) from the rotation axis of the disc, in particular the early Universe, seem to support the formation of stellar particles in counter-rotating orbits, as we can see from the increase in the mass of the counter-rotating component (which is increasing at a similar rate to the total stellar component) with redshift (left-hand panel). On the other hand, galaxies that do not show an increase in the mass of their counter-rotating component with redshift, had inflows of cold gas closely aligned with the rotation axis of the disc (right-hand panel).
It is possible that both properties (counter-rotating orbits and misaligned gas inflows) could be tracing some other mechanism, however, \cite{Jimenez2022} found that the vertical velocity dispersion of cold gas correlates most strongly with the specific gas accretion rate onto the disc as well as with the degree of misalignment between the inflowing gas and the disc’s rotation axis. This evidence supports misaligned accretion to be a main driver of gas kinematics.

It would be interesting to check if similar results can be confirmed by observations. Counter-rotating components formed at high redshift can be distinguished from those formed through mergers by looking at the ages of the components. However, ages cannot help us distinguish between counter-rotation due to mergers and due to misaligned inflows, since in both cases we would have counter-rotating components with relatively younger ages than the components formed at high redshift. One possible way to determine their formation could be to study the metallicity of these counter-rotating components, or to look at the globular cluster population in the galaxy: metal-rich (red) clusters are born from an initial central burst, while metal-poor (blue) clusters are brought in by the later accretion of less massive satellites \citep{Searle1978,Beasley2002,Tonini2013, Leaman2013, Kruijssen2019, Beasley2018}. If the galaxy had no mergers, most of its globular clusters would be red, metal-rich globular clusters. If the galaxy had a merger, there should be evidence of the presence of blue, metal-poor globular clusters. 

\section{Conclusions}
We analysed the orbital distributions of galaxies at $z \sim 0$ in the {\sc Eagle} simulations in order to shed light on the connection between their orbital components and their evolution history. We divided each galaxy into four different components, depending on their orbital circularity: a cold component (with close to circular orbits), a warm component, a hot component, and a counter-rotating component. We calculated the fraction of stellar particles and their median age in each component. We find that:
\begin{itemize}
\item Changes in the orbital distributions of galaxies are correlated with the stellar mass of the galaxies, so that galaxies have higher fractions of hot and counter-rotating orbits and lower fractions of cold and warm orbits with increasing stellar mass (Fig. \ref{fig:orbits_mass28}). We also find small differences when considering galaxies that did not experience any minor and/or major merger since $z \sim2$ and galaxies that have undergone minor and/or major mergers.
\item Stellar particles are older with increasing stellar mass regardless of their orbital family. Stellar particles on cold orbits are generally the youngest, with the oldest particles being on counter-rotating or hot orbits, at all masses. We also find small differences depending on whether galaxies had mergers or not. Stellar particles on cold orbits in galaxies with no mergers are slightly younger than those in galaxies with mergers, while the opposite is seen for counter-rotating orbits, where we see older stellar particles in galaxies with no mergers compared to those with mergers.  $\sigma$~Age of the cold component shows no dependence on stellar mass, while the warm, hot, and counter-rotating components have lower values of $\sigma$ Age with increasing stellar mass (Fig. \ref{fig:age_sigma_mass28_re}).
\item Galaxies with $M_{\star}/\rm M_{\odot}>10^{10.5}$ at $z=0$, have a peak in the fraction of cold and warm orbits at $z \sim 1$, with the most massive galaxies reaching their peak earlier ($z \sim 2$), and then steadily decrease with time. Opposite trends with time are found for hot and counter-rotating orbits, so that the orbital fractions are higher at $z=0$ for high-mass galaxies. For galaxies with $M_{\star}/\rm M_{\odot}<10^{10.5}$ at $z=0$, the fraction of cold and warm orbits steadily increases from $z=2$ to $z=0$, albeit with little evolution between $z=0$ and $z=0.5$. Hot and counter-rotating orbits, on the other hand, decrease with decreasing redshift. The ages of the components increase with increasing redshift. $\sigma$ Age decreases going towards high masses for all but the cold orbits, where it stays the same. (Fig. \ref{fig:orbits_redshift}).
\item The fractions of orbits are closely connected to the number of mergers a galaxy experiences. In particular, dry mergers seem to be the main channel to reduce the contributions from warm and cold orbits and to increase the contributions from hot and counter-rotating orbits (Fig. \ref{fig:all_pars_nmergers_mid} and Fig. \ref{fig:orbs_wetdry}).
\item At fixed stellar mass, for stellar masses $\log M_{\star}/ \rm M_{\odot} > 10.5$, galaxies with higher fractions of hot and counter-rotating orbits show higher fractions of ex-situ particles (Fig. \ref{fig:exsitu}).
\item Counter-rotating components have three possible channels for their formation: 
\begin{enumerate}
\item at high redshift, during a phase where both prograde and retrograde mergers contribute nearly equally to galaxy bulge growth;
\item through mergers, in particular dry mergers (Fig. \ref{fig:orbs_wetdry}).
\item in the absence of mergers, inflows with inclination angles highly misaligned from the rotation axis of the disc, can lead to an increase in the fractions of counter-rotating orbits (Fig. \ref{fig:fcr_angle}).
\end{enumerate}
\end{itemize}

\section*{Data Availability}

The data underlying this article are available through \href{http://icc.dur.ac.uk/Eagle}{http://icc.dur.ac.uk/Eagle}. Data products (orbital distributions) created for this article are available upon request to the \sendemail{giulia.santucci@uwa.edu.au}{Eagle orbital distributions: data request}{corresponding author}.

Measurements from Schwarzschild models of SAMI galaxies are available
\sendemail{giulia.santucci@uwa.edu.au}{SAMI Schwarzschild models: data request}{
contacting the corresponding author.}

\section*{Acknowledgements}

GS and KH acknowledge funding from the Australian Research Council (ARC) Discovery Project DP210101945. GS acknowledges funding from the UWA Research Collaboration Awards 2022. CL has received funding from the ARC Centre of Excellence for All Sky Astrophysics in 3 Dimensions (ASTRO 3D) through project number CE170100013.
CF is the recipient of an Australian Research Council Future Fellowship (project number FT210100168) funded by the Australian Government.
CL and CF are the recipients of ARC Discovery Project DP210101945.
RMcD acknowledges funding support via an Australian Research Council Future Fellowship (project number FT150100333).
AP is supported by the Science and Technology Facilities Council through the Durham Astronomy Consolidated Grant 2020–2023 (ST/T000244/1).
KLP acknowledges support from the Australian Government Research Training Program Scholarship.
GvdV acknowledges funding from the European Research Council (ERC) under the European Union's Horizon 2020 research and innovation programme under grant agreement No 724857 (Consolidator Grant ArcheoDyn).



\bibliographystyle{mnras}
\bibliography{references} 




\appendix
\section{{\sc Eagle} High-resolution dark matter component}\label{app:hrdm}
\begin{figure*}
\centering
\includegraphics[scale=0.49, trim=0cm 0cm 0cm 0cm, clip=True]{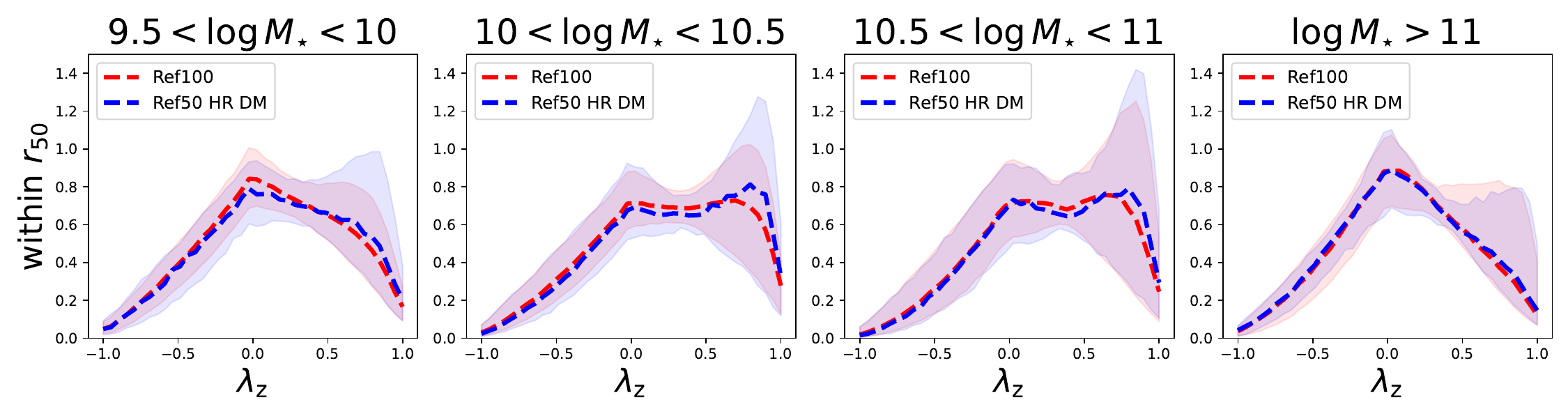}
\caption{Distribution of $\lambda_z$ for four different stellar mass bins calculated in the Reference run (red) and in the high-resolution dark matter run (blue). In general, the distributions for the two runs are very similar, with only small differences for stellar masses 10$< \log M_{\star} / M_{\odot} <$10.5 and in the scatter of the cold orbits. }
\label{fig:lambda_z_high_res_dm}
\end{figure*}
We show in Fig. \ref{fig:lambda_z_high_res_dm} the distribution of $\lambda_z$ considering the total galaxy for the reference run (Ref100, in red) and the high-resolution dark matter run (Ref50 HR DM, in blue), for different mass bins. In general, the distributions for the two runs are very similar, with only small differences for stellar masses 10$< \log M_{\star} / M_{\odot} <$10.5 and in the scatter of the cold orbits. 

If we calculate the fraction of orbits for both runs, we find good agreement between the median values of all the components. Note that the stellar mass used for this figure is the total subhalo stellar mass, not the stellar mass within 30kpc used in Fig. \ref{fig:orbits_mass28}.
\begin{figure}
\centering
\includegraphics[scale=0.5, clip=True]{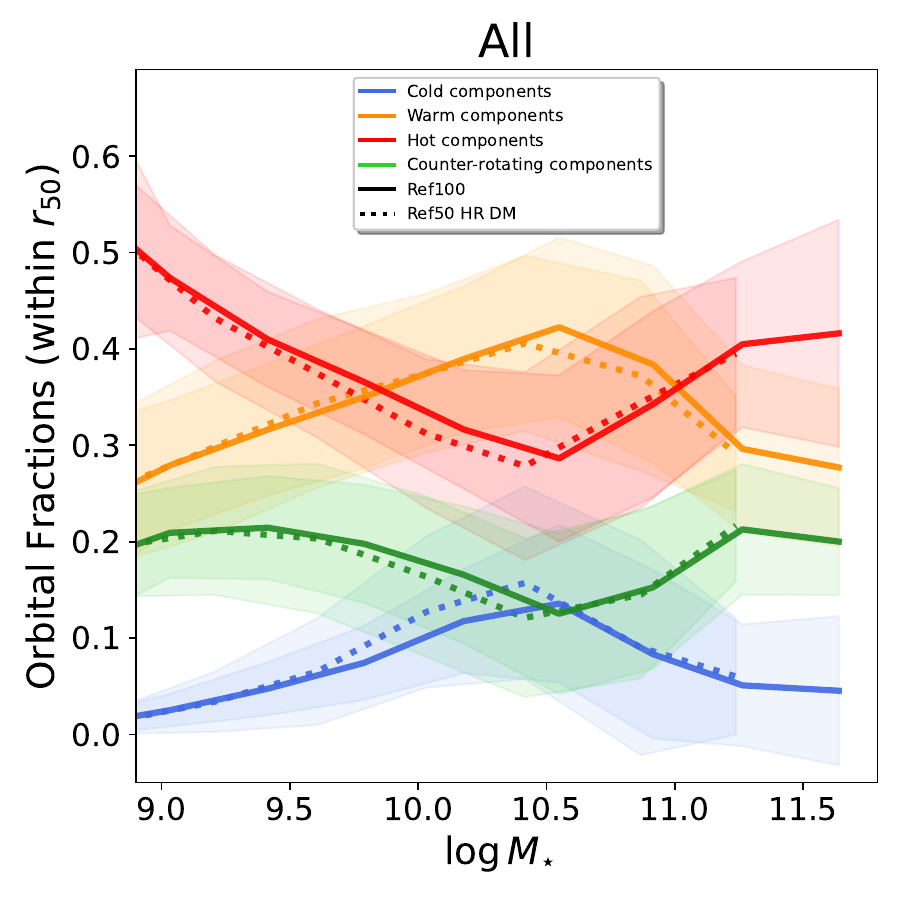}
\caption{Fraction of orbits (mass-weighted) at $z \sim 0$ as a function of the total stellar mass in the {\sc Eagle} subhaloes. Solid lines show the median values (for bins with at least 10 galaxies) and the shaded area delimits the 1-$\sigma$ scatter. Cold orbits are shown in blue, warm orbits in orange, hot orbits in red, and counter-rotating orbits in green, as labelled.}
\label{fig:fracs_high_res_dm_all}
\end{figure}

\section{Ages and $\sigma$~Age of the total orbital components}\label{app:total_age}
In this Section, we show the median values of Age and $\sigma$~Age for the orbits within the whole galaxy as a function of stellar mass. We find that median Ages and  $\sigma$~Age of the different components in the whole galaxy show very similar trends to those within 1 $r_{\rm 50}$. 
\begin{figure*}
\centering
\includegraphics[scale=0.45 ,clip=True]{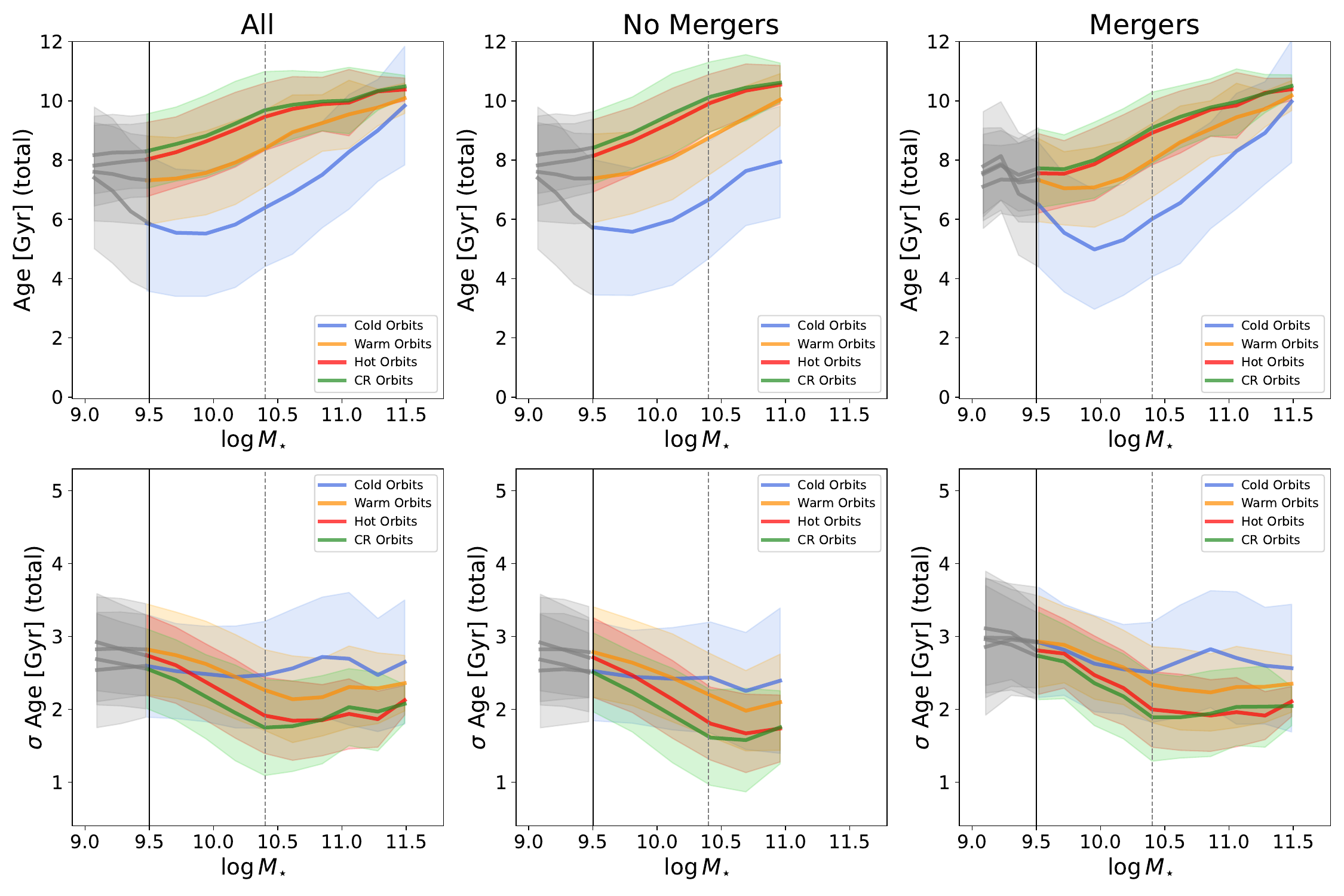}
\caption{Top panels: Median age of stellar particles in each total component at $z \sim 0$ as a function of stellar mass. Bottom panels: Median $\sigma$~Age of stellar particles in each total orbital component at $z \sim 0$ as a function of stellar mass. Lines and panels are as in Fig.~\ref{fig:orbits_mass28}. All ages increase with increasing stellar mass. Particles on cold orbits are generally younger than particles on warm orbits, with the oldest particles being on hot or counter-rotating orbits. Stellar particles on cold orbits in galaxies with no mergers are younger than those in galaxies with mergers; the opposite is seen for counter-rotating orbits, where we see older stellar particles in galaxies with no mergers compared to those with mergers. We find that $\sigma$~Age of the cold component shows no dependence on stellar mass, while the warm, hot, and counter-rotating components have a lower scatter in the ages of their stellar particles with increasing stellar mass. Particles on cold orbits have generally a wider diversity in ages than particles on warm orbits. Particles in hot and contour-rotating orbits have the lowest spread in ages.}
\label{fig:age_sigma_mass28_all}
\end{figure*}

\section{Stellar orbital families of {\sc Eagle} galaxies across cosmic time, normalised by the age of the Universe}\label{app:hubbletime}
\begin{figure*}
\centering
\includegraphics[scale=0.33 ,trim=4cm 6cm 4cm 6cm, clip=True]{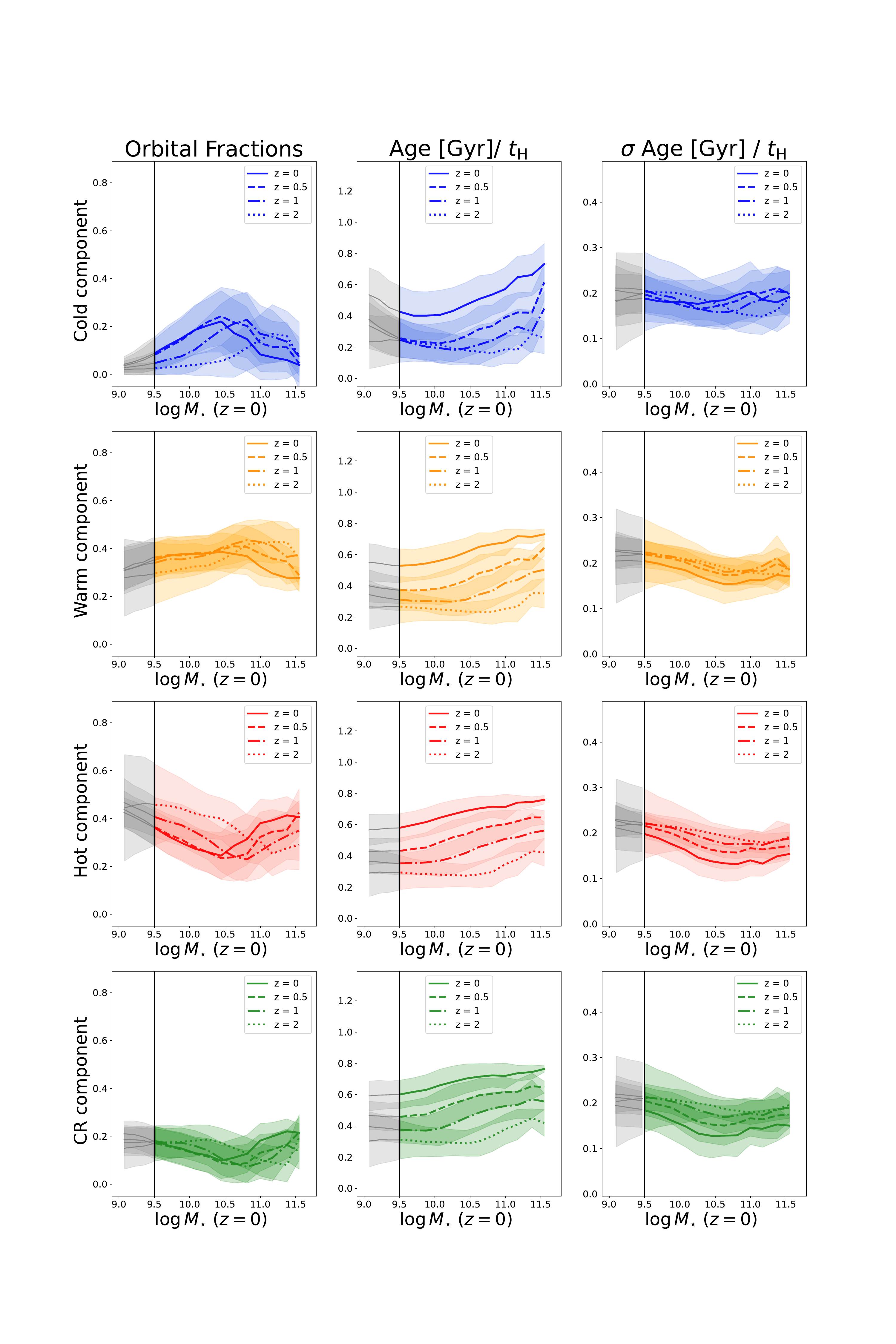}
\caption{Orbital fractions (panels on the left), stellar Age (middle panels), and $\sigma$~Age (right-hand panels) as a function of the $z \sim 0$ galaxy stellar mass at different redshifts, as labelled in each panel. Cold orbits are shown in blue (top panels), warm orbits in orange (second from top panels), hot orbits in red (second from bottom panels), and counter-rotating orbits in green (bottom panels). We show the median values for each component at different redshift: $z \sim 0$ as a solid line, $z \sim 0.5$ as a dashed line, $z \sim 1$ as a dash-dotted line, and $z \sim 2$ as a dotted line. Shaded regions indicate the 1~$\sigma$ scatter from the median. For all 4 snapshots, we see the fractions of orbits increasing with increasing stellar mass till they reach a transition mass and then they start to decrease. This transition mass changes with redshifts. The ages of the components increase linearly with increasing stellar mass. The degree of increase changes with redshift, becoming steeper at lower redshifts. The scatter in the ages decreases going towards high masses for all but the cold orbits, where it stays the same.}
\label{fig:htime}
\end{figure*}

We show the fraction of orbits, age, and $\sigma$ Age of the stellar particles in each component as a function of stellar mass at $z=0$, $z=0.5$, $z=1$, and $z=2$ in Fig.~\ref{fig:htime}. Ages and $\sigma$ Age of each component are normalised by the Hubble time ($t_{\rm H}$) at each redshift. The trends we see here are similar to those found in Fig. \ref{fig:orbits_redshift}, with different absolute values due to the normalisation, so that at fixed redshift the ages of the components increase with increasing stellar mass. The degree of increase with stellar mass changes with redshift, becoming steeper at lower redshifts. Similarly, at fixed stellar mass, the ages increase with decreasing redshift. $\sigma$ Age decreases going towards high masses for all but the cold orbits, where it stays the same.

\section{Orbital fractions and number of mergers}\label{app:mergers}
\begin{figure*}
\centering
\includegraphics[scale=0.35 , clip=True]{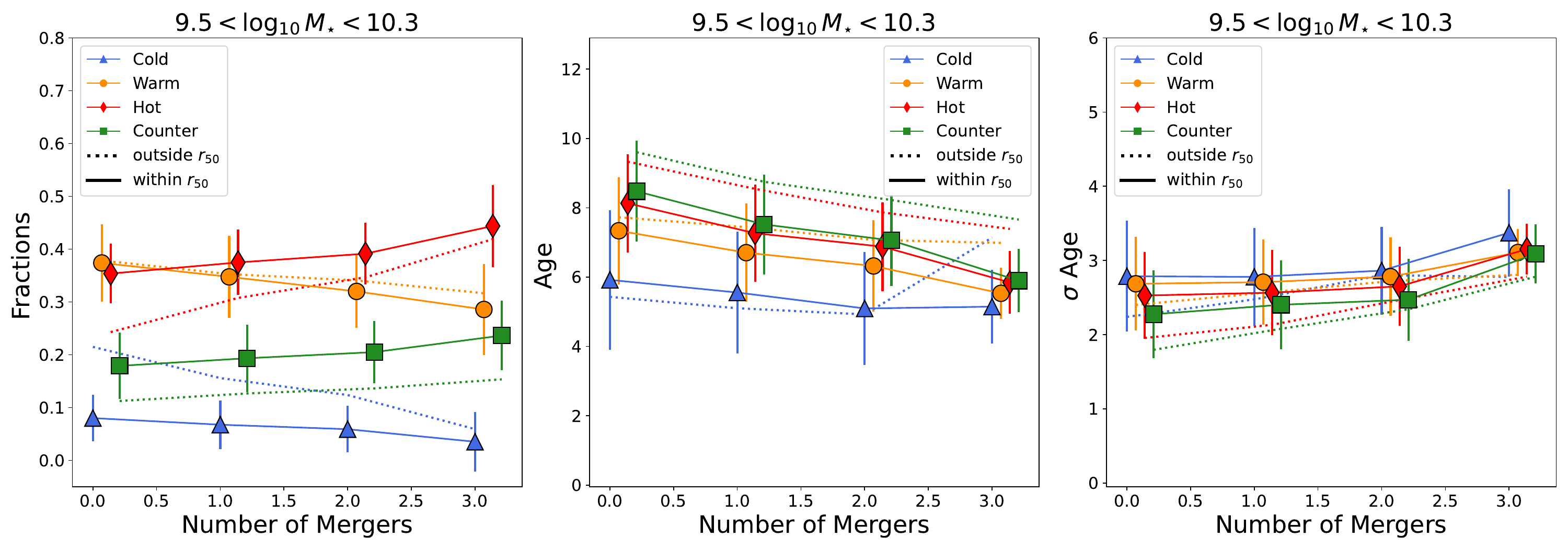}
\caption{{\it Left-hand panel:} Medians (symbols connected by lines) of the stellar mass fractions in each orbital component, as labelled, within 1~$r_{\rm 50}$ (solid lines) and outside 1~$r_{\rm 50}$ (dotted lines) for galaxies $ 9.5 < \log M_{\star}/ \rm M_{\odot} < 10.3$ at $z=0$, as a function of the total number of minor plus major mergers experienced by the galaxies. The latter is the cumulative number of merger events at $0\le z\le 2$. Errorbars delimit the 1-$\sigma$ scatter around the median. The stellar mass fraction in warm and cold orbits decreases with increasing number of mergers, while the fractions of hot and counter-rotating orbits increase with increasing number of mergers. 
{\it Middle panel:} As in the left-hand panel but for the mean stellar age of each orbital component. 
Particles in galaxies that have undergone a higher number of mergers have younger ages, independently of the component they belong to, with the cold component having the youngest ages overall. 
{\it Right-hand panel:} As is the left-hand panel but for $\sigma$~Age of each orbital component. There is an increase in $\sigma$~Age for the stellar particles in each orbital component with increasing numbers of mergers.}
\label{fig:all_pars_nmergers_low}
\end{figure*}

\begin{figure*}
\centering
\includegraphics[scale=0.35 , clip=True]{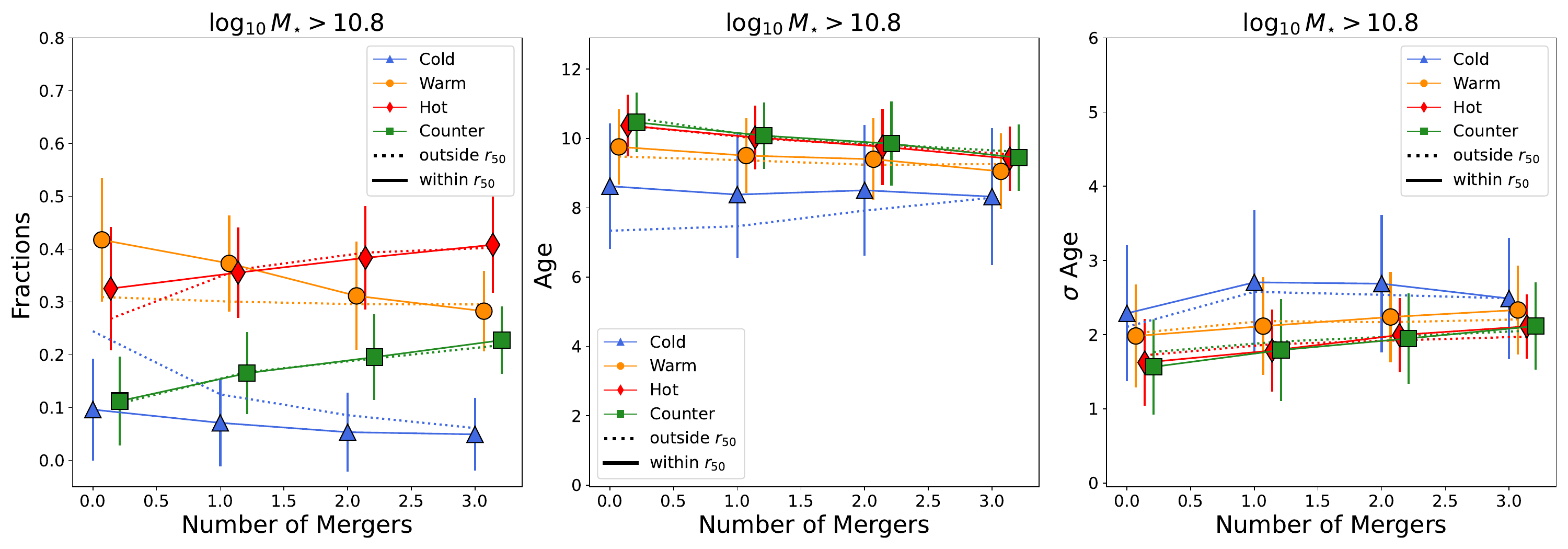}
\caption{Lines and panels as in Fig. \ref{fig:all_pars_nmergers_low} for galaxies with $\log M_{\star} > 10.8$. The stellar mass fraction in warm and cold orbits decreases with increasing number of mergers, while the fractions of hot and counter-rotating orbits increase with increasing number of mergers. Particles in galaxies that have undergone a higher number of mergers have younger ages, independently of the component they belong to, with the cold component having the youngest ages overall. There is an increase in $\sigma$~Age for the stellar particles in each orbital component with increasing numbers of mergers. }
\label{fig:all_pars_nmergers_high}
\end{figure*}

In this Section, we show the median values of the orbital fractions within and outside 1 $r_{\rm 50}$ as a function of the number of minor and major mergers each galaxy has undergone, for galaxies with low mass ($ 9.5 < \log M_{\star}/ M_{\odot} < 10.3$; Fig. \ref{fig:all_pars_nmergers_low}), and high mass ($\log M_{\star}/ M_{\odot} < 10.8$; Fig. \ref{fig:all_pars_nmergers_high}). Low-mass galaxies and high-mass galaxies show similar global trends to those of galaxies with intermediate mass (shown in Fig. \ref{fig:all_pars_nmergers_mid}), although with different absolute values. 

\section{Ex-situ fractions within 1 $r_{\rm 50}$}\label{appendix:exsitu_re}

\begin{figure*}
\centering
\includegraphics[scale=0.4 , clip=True]{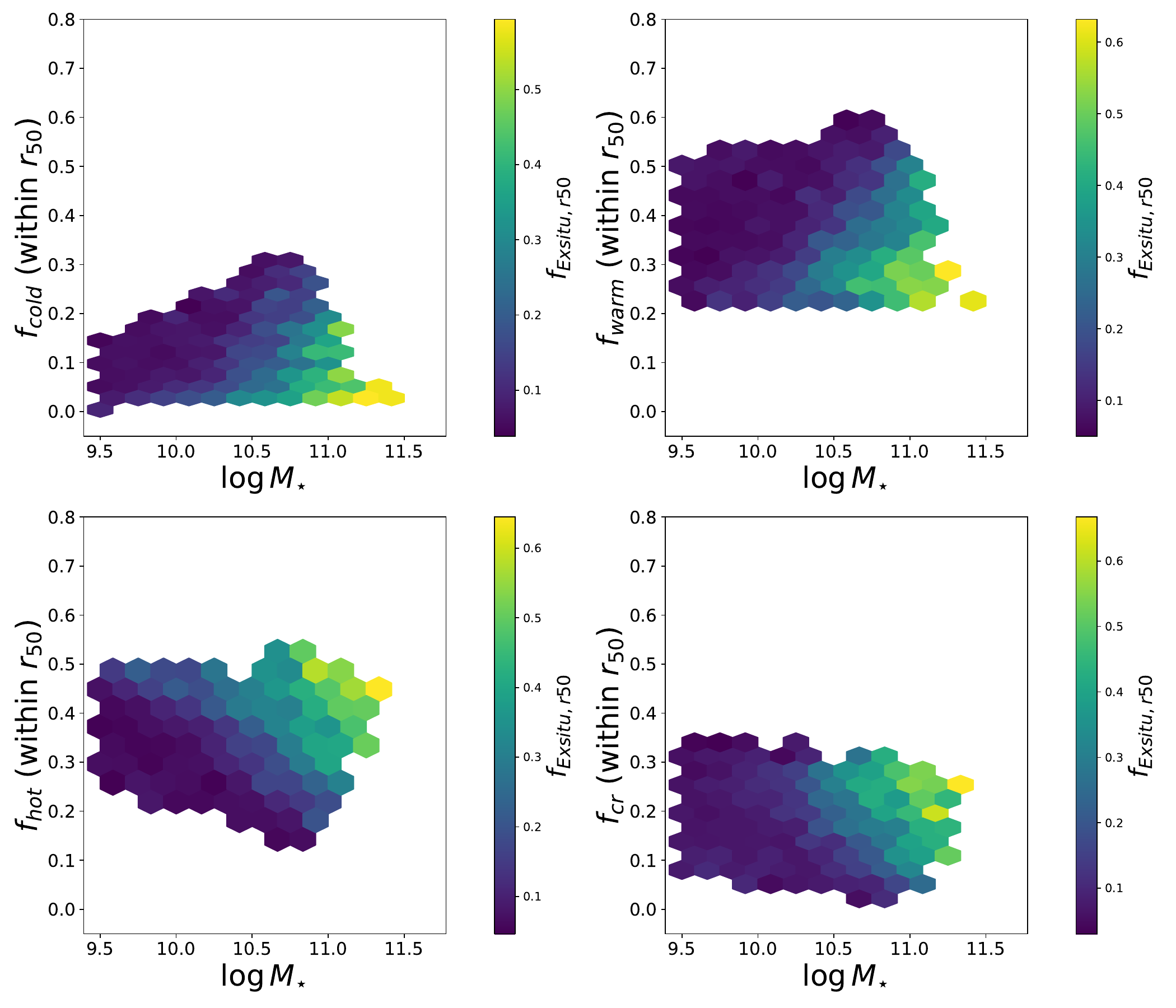}
\caption{Fraction of orbits in different components within 1 $r_{\rm 50}$ as a function of stellar mass, colour-coded by their total ex-situ fractions. Each hexabin includes at least 5 galaxies and is colour-coded by the mean of the total ex-situ fractions for the galaxies included in the bin. Low-mass galaxies have very little contribution from ex-situ fractions, which have a sharp increase around $ \log M_{\star}/ \rm M_{\odot} \sim 10.5$. For stellar masses above $ \log M_{\star}/ \rm M_{\odot} \sim 11$, at fixed stellar mass, galaxies with higher ex-situ fractions have higher fractions of hot and counter-rotating orbits and lower fractions of cold and warm orbits.}
\label{fig:exsitu_re}
\end{figure*}

We show in Fig. \ref{fig:exsitu_re} the fraction of orbits in different components within 1 $r_{\rm 50}$ as a function of stellar mass, colour-coded by their total ex-situ fractions. Low-mass galaxies have very little contribution from ex-situ fractions, which have a sharp increase above $ \log M_{\star}/ M_{\odot} \sim 10.5 - 10.6$. These trends are consistent with what we see when considering the whole galaxy (Fig. \ref{fig:exsitu}), although less strong.


\bsp	
\label{lastpage}
\end{document}